\documentclass{aastex631}
\usepackage{subfigure}
\usepackage{multirow}
\usepackage{longtable}
\usepackage{amsthm,amsmath,amssymb}
\usepackage{mathrsfs}

\shorttitle{A comparative analysis of two peculiar long Gamma-ray bursts: GRB 230307A and GRB 211211A }

\shortauthors{Peng et al}
\graphicspath{{./}{figures/}}

\begin{document}

\title{A comparative analysis of two peculiar Gamma-ray bursts: GRB 230307A and GRB 211211A}

\author{Zhao-Yang Peng}
\affiliation{College of Physics and Electronics Information, Yunnan Normal University, Kunming 650500, China \\}
\author{Jia-Ming Chen}
\affiliation{School of Physics and Astronomy, Yunnan University, Kunming 650500, China\\}
\author{Jirong,Mao}
\affiliation{
Yunnan Observatories, Chinese Academy of Sciences, Kunming 650011, China\\}
\affiliation{Center for Astronomical Mega-Science, Chinese Academy of Sciences, Beijing 100101, China\\}
\affiliation{Key Laboratory for the Structure and Evolution of Celestial Objects, Chinese Academy of Sciences, Kunming 650011, China}

\begin{abstract}
GRB 211211A is a peculiar long Gamma-ray burst (GRB) with very high brightness and short burst properties. It's full lightcurve consists of three emission episodes, i.e. a precursor, a main burst and a extended emission. We find a recently detected long-duration GRB 230307A also includes the three consistent emission episodes. Furthermore, the two bursts have similar redshift 0.076 and 0.065, respectively. We perform a detail temporal and spectral analysis of the two GRBs to compare their temporal and spectral properties. Our analysis shows that the two bursts share great similarities for both the whole emission and the three corresponding emission phases, which are listed as follows: (1) they have near zero spectral lag, (2) they have very short minimum variability timescale (MVT),  (3) they lie in the same region of in the MVT-$T_{90}$, Amati relation, and hardness-$T_{90}$ planes, (4) the three phases are quasi-thermal spectra, (5) both the peak energy and the low-energy index track the flux, (6) the time-resolved spectra are much wider than those of the blackbody prediced by theory model, (7) there are strong correlations between thermal flux and total flux and the correlation coefficients as well as the slopes for the corresponding stages are very consistent, (8) the photosphere emission properties are very consistent. Other investigations and observations suggest the two GRBs indeed belong to short burst with a compact star merger origin. Therefore, we think that GRB 230307A and GRB 211211A are the rare and similar GRBs and the photospheric radiation can interpret their radiation mechanisms.

\end{abstract}

\keywords{Gamma-ray bursts(629)}

\section{Introduction} \label{sec:intro}
Although more than fifty years have passed since the discovery of gamma-ray bursts (GRBs), the classification, the radiation mechanism of the prompt emission, and the jet composition in GRBs are still three unsolved issues \citep{2018pgrb.book.....Z}. The earliest classification of the GRBs was put forward by the bimodality distribution based on the duration, $T_{90}$, of GRBs observed by the BATSE instrument of CGRO. An increasing number of the GRBs detected by several instruments on board some satellites, such as BeppoSax, Fermi-GBM, Swift, and Konus-Wind have confirmed the classification. Generally speaking, it is believed that long-duration GRBs (LGRBs, also Type II) are associated with the massive star core-collapse \citep{1993ApJ...405..273W, 1998ApJ...494L..45P,2006ARA&A..44..507W,2006RPPh...69.2259M}. Many observed features were accounted for within the collapsel model, for example, the associations with Type Ic supernovae (SNe) \citep{1998Natur.395..670G,1998Natur.395..663K,2017A&A...605A.107C}. Whereas short-duration GRBs (SGRBs, also Type I) originate from the mergers of compact star binaries and a lot of evidences, such as the discovery of GRB 170817A and GW 170817, confirmed that binary neutron star (BNS) mergers do produce SGRBs.  Some observations suggest that kilonovae (KNe) have been observed in association with at least some SGRBs \citep{1986ApJ...308L..43P,1989Natur.340..126E,1992ApJ...397..570M,2017ApJ...848L..13A,2017ApJ...848L..14G}. Therefore, it is believed that SGRBs are often associated with KNe.

However, more and more observations revealed that such a classification does not always hold. For example, GRB 060505 and GRB 060614 are two long-duration GRBs but no supernovae was observed to be associated with them \citep{2006Natur.444.1047F}. Moreover, the properties of GRB 060614 have some similar features with those of SGRBs \citep{2006Natur.444.1044G}. GRB 211227A with a $T_{90}$ of about 84 s also has no detected associated supernova signature after performed follow-up observations of the optical emission \citep{2022ApJ...931L..23L}. However, it is very interesting that GRBs 060614 and 211227A show evidence of accompanying KNe \citep{2015NatCo...6.7323Y}.

Some arguments such as \cite{2009ApJ...703.1696Z}, \cite{2011ApJ...727..109V}, and \cite{2012ApJ...749..110B}, have suggested that at least some SGRBs are probably not related to compact star mergers. For instance, an observer-frame short GRB 090426 shared many properties of long GRBs with a massive star origin and some correlations are similar to those seen in LGRB \citep{2009A&A...507L..45A,2010MNRAS.401..963L,2011MNRAS.410...27X}. Another short burst GRB 200826A, with a rest-frame duration of $\sim$0.5 s, is consistent with LGRBs because this event was energetic and soft and had associated an SN \citep{2022ApJ...932....1R}. 

A nearby long burst GRB 211211A with peculiar origin has been detected recently. Several investigations consistently showed that GRB 211211A originated from a compact merger, and a kilonova emission was discovered in the optical/NIR band in temporal and spacial coincidence with the burst \citep{2022Natur.612..223R,2022arXiv220610710W,2023NatAs...7...67G,2022Natur.612..228T,2022ApJ...936L..10Z}. Therefore, GRB 211211A further demonstrates that mergers may drive LGRBs. \cite{2022Natur.612..232Y} suggested that a white dwarf (WD)-NS merger with a post-merger magnetar engine provides a self-consistent interpretation for the prompt emission, the X-ray afterglow emission, and the engine-fed kilonova emission. \cite{2023ApJ...947L..21Z} further showed that GRB 211211A associated kilonova-like emission may arise from a neutron star-white dwarf (NS–WD) merger if the central engine leaves a magnetar after the burst.

As pointed out by \cite{2022Natur.612..236M}, the GeV emission of GRB 211211A is the first ever high-energy component observed with a compact binary merger event. No Fermi/LAT detection was reported on timescales of minutes, hours, and days after the BNS merger for GW170817. Moreover, similar to the three-stage GRB 160625B, the full lightcurve of GRB 211211A also can be divided into three emission episodes: a precursor with duration of  $\sim$ 0.2 s, a spiky hard main emission of $\sim$ 8 s, and a soft long extended emission up to $>$ $\sim$ 50 s. This burst has a duration of $T_{90}$ $\sim$ 34.3 s in 10-1000 keV as detected by Fermi-GBM. Different from GRB 160625B,  the quiescent period between the precursor and the main emission of GRB 211211A ($\sim$1 s) is much shorter than that of GRB 160625B (180 s) \citep{2022arXiv220502186X,2018NatAs...2...69Z}. It is interestingly noticed that LGRB 230307A seems to have similar three-stage lightcurve to GRB 211211A and it represents one of the brightest events detected in over 50 years of observation \citep{2022cxo..prop.6493F}. More interesting thing about GRB 230307A is reported by \cite{2023GCN.33569....1L} that the burst may arise from a compact binary ejected from the nearby galaxy. Other researchers such as \cite{2023GCN.33578....1B} and \cite{2023GCN.33747....1L} also support the claim of kilonova emission given by \cite{2023GCN.33569....1L}. 

Since GRB 230307A and GRB 211211A have similar three-stage lightcurves and are LGRBs with similar KNe origin, we would like to compare them based on thetemporal and spectral properties in order to investigate their unknown characteristics of GRB 230307A and GRB 211211A. The paper is organized as follows: In Section \ref{Date and methods}, we present the observations and data analysis of the GRB 211211A and GRB 230307A. In Section \ref{results}, we introduce the temporal and spectral analysis results of two GRBs. In Section \ref{photosphere emission}, we estimate the physical parameters of the outflow from the photosphere radiation. In Section \ref{Summary}, we summarize our results with some discussion.

\section{Observations and data analysis\label{Date and methods}}
Both GRBs triggered recently the Fermi Gamma-ray Burst Monitor (GBM). GRB 211211A (trigger 660921004/211211549)  triggered the Fermi-GBM at 13:09:59.651 UT on 11 December 2021, and a duration of $T_{90}\sim34.3 s$ in the $50-300 keV$ band was estimated \citep{2021GCN.31210....1M}.  The redshift reported by \citet{2021GCN.31221....1M} is 0.076. GRB 230307A (trigger 699896651/230307656) triggered the Fermi-GBM at 15:44:06.67 UT on 7 March 2023, and a duration of $T_{90}\sim39.88 s$ in the $50-300 keV$ band was estimated \citep{2023arXiv230311083W}. The redshift reported by \citet{2023GCN.33485....1G} is 0.065. This event is one of the brightest GRBs among the GBM detections. 

We further analyze the spectral properties of GRB 211211A and GRB 230307A observed by Fermi-GBM. The GBM has 12 sodium iodide (NaI) scintillation detectors covering the 8 keV-1 MeV energy band, and two bismuth germanate (BGO) scintillation detectors sensitive to the 200 keV-40 MeV energy band \citep{2009ApJ...702..791M}. In principle, the selection of detectors to perform spectral analysis is based on the count rate and pointing direction. We should also consider some special cases. For GRB 230307A, \cite{2023GCN.33551....1D} reported that the bad time interval due to the pulse pile-up is T0+[3.00, 7.00] s (T0 is the GBM trigger time) for the TTE data. Thus we select one NaI (na) detector and one BGO (b1) detector for data analysis, and the bad time interval is marked in some Figures by the shaded region. For GRB 211211A, there are three NaI detectors with the pointing angle less than $60^{\circ}$ named n1, n2, and na. The most illuminated BGO detector is b0. Therefore, detectors n1, n2, na, and b0 are used for our spectral analysis.

We adopt the Bayesian analysis package, the multi-task maximum likelihood framework (3ML; \cite{2021zndo...5646953B}), to perform time-integrated and time-resolved spectral analysis of the two GRBs. We
have successfully used it for the data analysis in several previous works (e.g. \cite{2022ApJ...940...48D,2022ApJ...932...25C,2021ApJ...922...34C,2021ApJ...920...53C}). This fully Bayesian approach is widely used in the Fermi-GBM catalogue for bright GRB spectra \citep{2019ApJ...886...20Y,2019ApJS..245....7L}. The background fitting is performed by the brightest NaI detector in photon counting, and two typical off-source (pre- and post source) intervals are selected for the fitting with a polynomial of order $0 - 4$, and the optimal order of the polynomial tests are determined by likelihood ratios. This optimal polynomial is then applied to fit each of the 128 energy channels. To perform time-resolved spectral analysis, the light curve must be re-divided into sufficient intervals. We use a Bayesian block method to restructure the TTE light curves of the brightest NaI detectors with a false alarm probability of $p = 0.01$. Similar to \citet{2018ApJS..236...17V}, we adopt $S > 10$ as the criterion for selecting time bins containing enough source photons to obtain good fitting results.

The energy spectrum of a GRB can usually be well described by a smoothly connected power-law function (``Band" function; \cite{1993ApJ...413..281B}). The Band function defined as
\begin{equation}
N{{\left( E \right)}_{Band}}=\left \{\begin{array}{clcc}
  & A{{\left( \frac{E}{100keV} \right)}^{\alpha }}\exp \left( -\frac{E}{{{E}_{0}}} \right),E<\left( \alpha -\beta  \right){{E}_{0}} \\
 & A{{\left[ \frac{\left( \alpha -\beta  \right){{E}_{0}}}{100keV} \right]}^{\alpha -\beta }}\exp \left( \beta -\alpha  \right){{\left( \frac{E}{100keV} \right)}^{\beta }},E\ge \left( \alpha -\beta  \right){{E}_{0}} \\
\end{array}\right.
\end{equation}
where $A$ is the normalization constant in unit of $ph$  $cm^{-2}$ $keV^{-1}$ $s^{-1}$. $E_{0}$ is the break energy spectrum, $\alpha$ and $\beta$ are the low-energy and high-energy power-law spectral indices, respectively. The two spectral regimes are separated by $E_0$. The peak energy in the $E^2N(E)$ spectrum is related to $E_0$ by $E_p = (2+ \alpha)E_0$. Furthermore, the high-energy spectral index $\beta$ may not be well constrained if there are relatively small high-energy photon numbers. We can use the cutoff power-law function (CPL),
\begin{equation}
N(E)_{\textrm{CPL}}=A(\frac{E}{100\textrm{keV}})^{\alpha }\textrm{exp}\left [ -\frac{(\alpha+2)E}{E_{p}} \right ],
\end{equation}
where $A$ is the normalization factor in unit of the $ph$ $cm^{-2}$ $keV^{-1}$ $s^{-1}$. $E_p$ is the peak energy; $\alpha$ is the low energy power law photon spectral index.
Some bursts have an additional thermal component, typically fit by the Planck blackbody (BB) function, which is given by
\begin{equation}
    {{N(E)}_{BB}}=A\left( t \right)\frac{{{E}^{2}}}{\exp \left[ E/kT \right]-1},
\end{equation}
where $A$ is the normalization, $k$ is the Boltzmann constant, and $kT$ is the blackbody temperature.

In order to select the best model from the fixed models, we adopt the deviation information criterion (DIC). It is defined as
\begin{equation}
   DIC=-2\log \left[ p\left( data|\hat{\theta } \right) \right]+2{{p}_{DIC}},
\end{equation}
where $\hat{\theta }$ is the posterior mean of the model and $p_{DIC}$ is the effective number of parameters. The preferred model is the one with the lowest $DIC$ number.

\section{Analysis results\label{results}}
\subsection{The temporal profile properties}
\subsubsection{The temporal profile}
Here we compare the temporal profiles of the two bursts. We select NaI detector n1 for GRB 211211A and na for GRB 230307A to derive the GBM light curve by binning the photons with a bin size of 8 ms in the energy range of 8–900 keV. The temporal profiles of the two GRBs are shown in Figure 1. It is found that each light curve has a dim precursor and a very short quiescent time between the precursor and the main emission. The quiescent times are $\sim$ 0.30 s and $\sim$ 0.90 s for GRB 230307A and GRB 211211A, respectively. Dim extended emissions can be also found in GRB 211211A and GRB 230307A. The quiescent periods between the main and the extended emissions are $\sim$ 0.85 s and $\sim$ 2.00 s for GRB 230307A and GRB 211211A, respectively. The durations of the two precursors are $\sim$ 0.3 s and $\sim$ 0.2 s for GRB 230307A and GRB 211211A, respectively. Furthermore, the durations of the main emission and extended emission for the two bursts have significant differences: GRB 211211A exhibits a much shorter main emission phase but a twice longer extended emission phase in comparison with GRB 230307A (see Table 1).  Therefore, the three distinct phases, namely, precursor, main emission, and extended emission, are different as shown in the two GRBs. 

\subsubsection{The spectral lag}
Here we investigate the spectral lags of the three different phases for the two bursts in this subsection. A widely used cross-correlation function (CCF) is adopted to compute the GRB spectral lag \citep{1997ApJ...486..928B}. In this paper, we select a more suitable method for GRB temporal analysis to compute the spectral lags \citep{2015MNRAS.446.1129B}. This is a kind of modified CCF derived by \cite{1997ApJ...486..928B}. We also choose an asymmetric Gaussian model, because it provides the natural asymmetry of the CCF inherited by the asymmetry of GRB pulses \citep{1997ApJ...486..928B}. It has been tested by \citep{2015MNRAS.446.1129B}. The Markov Chain Monte Carlo (MCMC) method is used to fit the CCF. We show the results of our lag computation for the two GRBs with an asymmetric Gaussian function in Figure 2. The uncertainty of the lag has been estimated by applying a flux-randomization method that was presented by \citet{2015MNRAS.446.1129B}. 

The parameters of the spectral lags are listed in Table 1, and the CCFs with the time delay are shown Figure 2. It is found that the lags of corresponding episodes for the two bursts seem to be very consistent. Moreover, all lags are consistent with being zero, which makes both GRBs fall into the SGRB category.
\begin{figure}[ ]  
\centering
\begin{minipage}[t]{0.9\textwidth}
\centering
\includegraphics[width=\textwidth]{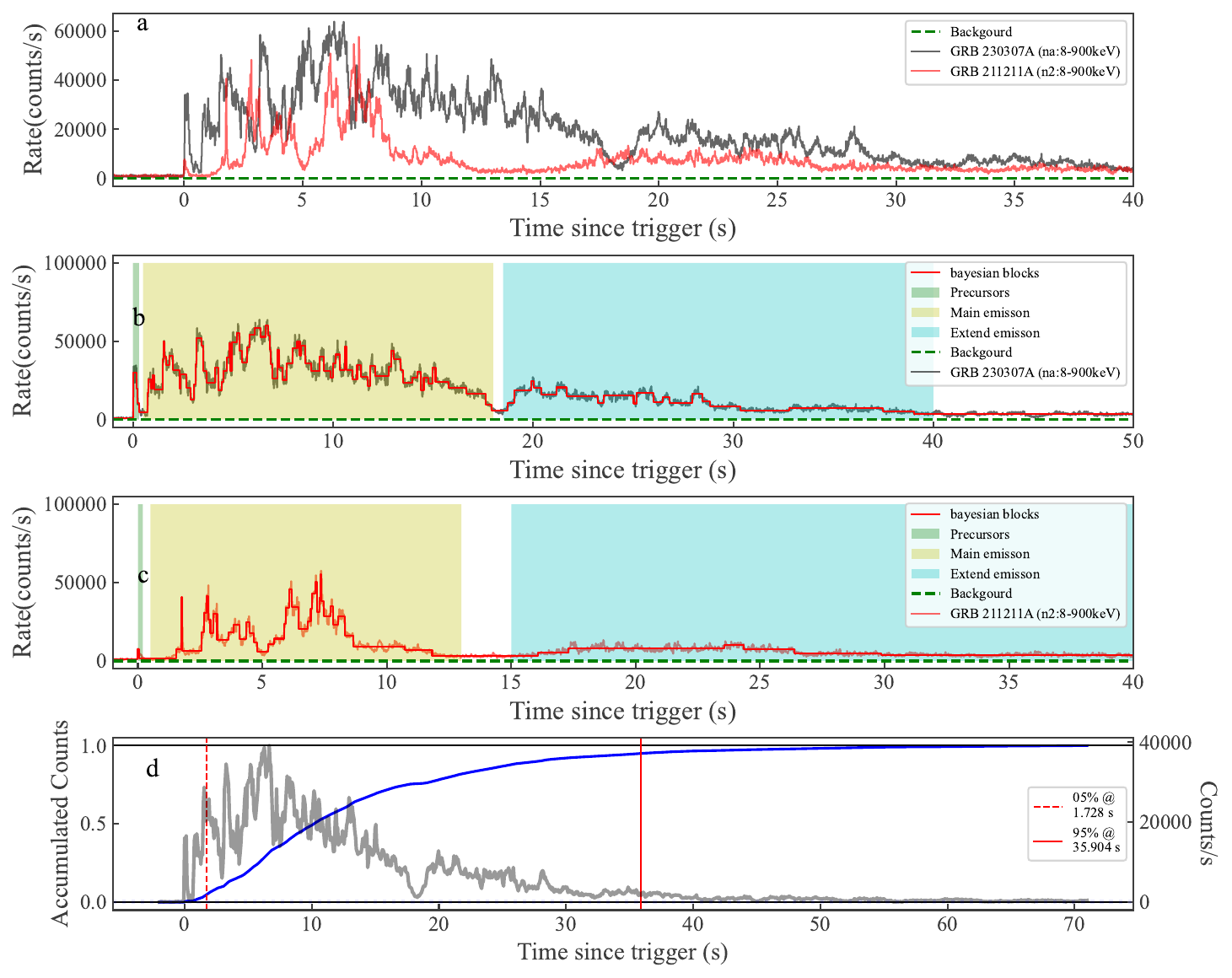}
\end{minipage}
\caption{The temporal behaviours of GRB 230307A and GRB 211211A.  a, The net light curves of GRB 230307A (black line) and  GRB 211211A (red line) obtained from Fermi-GBM data. b, The net light curve (blue line) of GRB 230307A obtained from Fermi-GBM data. The grey line is light curve and the red line is the Bayesian block, respectively.  The light cyan, light yellow and cyan shaded areas represent precursor, main emission and extended emission phases, respectively. c, The net light curve (red line) of GRB 211211A obtained from Fermi-GBM data. The lines and colors are the same as the panel c. d, The accumulated counts (blue line) of the Fermi-GBM net light curve. The grey horizontal dashed (solid) lines are drawn at 5\% (0\%) and 95\% (100\%) of the total accumulated counts. The $T_{90}$ interval is marked by the red vertical dashed (solid)  lines.  \label{fig2}}
\end{figure}

\begin{figure}[]
\centering
\begin{minipage}[t]{\textwidth}
\centering
\includegraphics[width=0.24\textwidth]{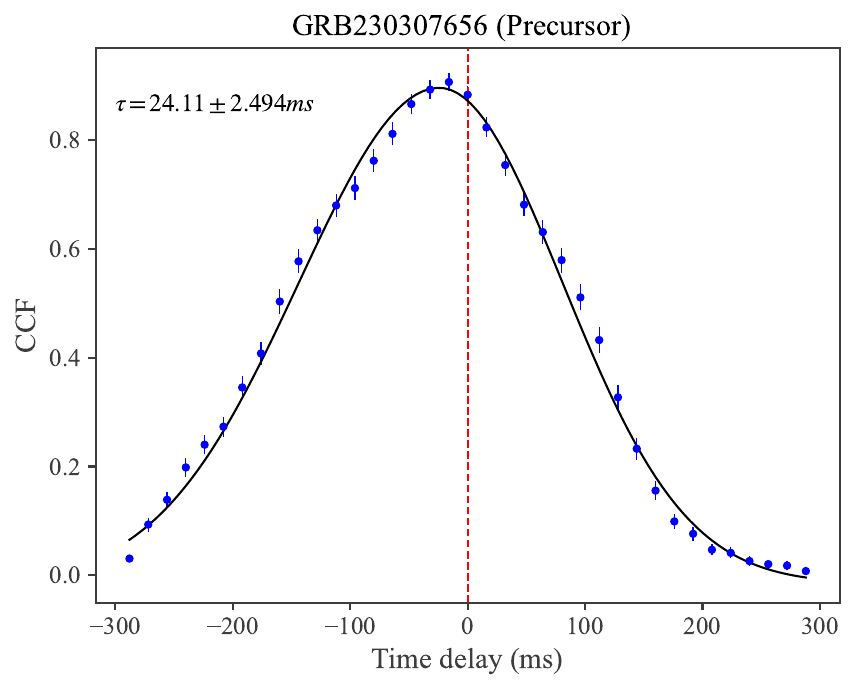}
\includegraphics[width=0.24\textwidth]{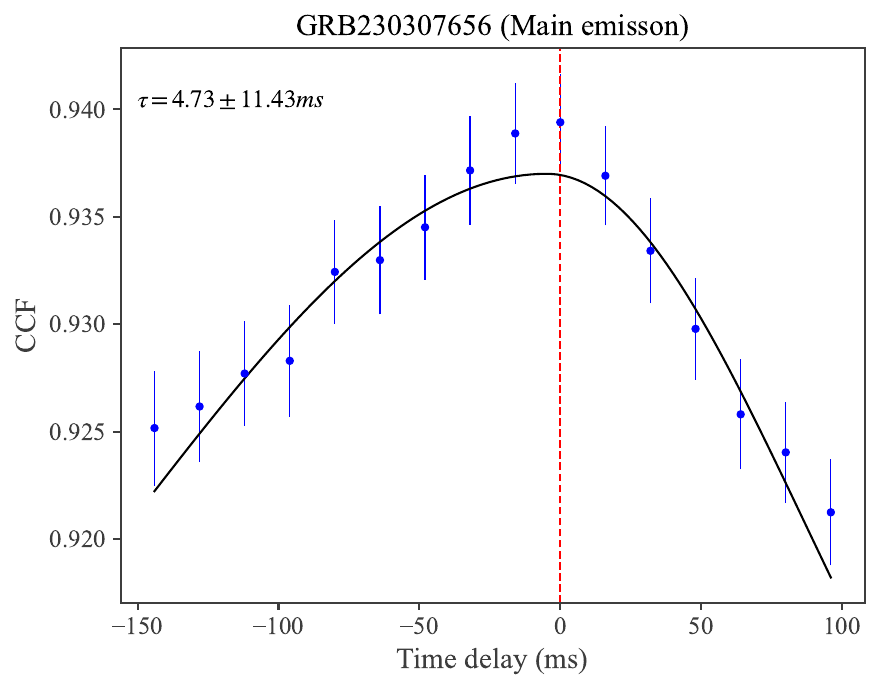}
\includegraphics[width=0.24\textwidth]{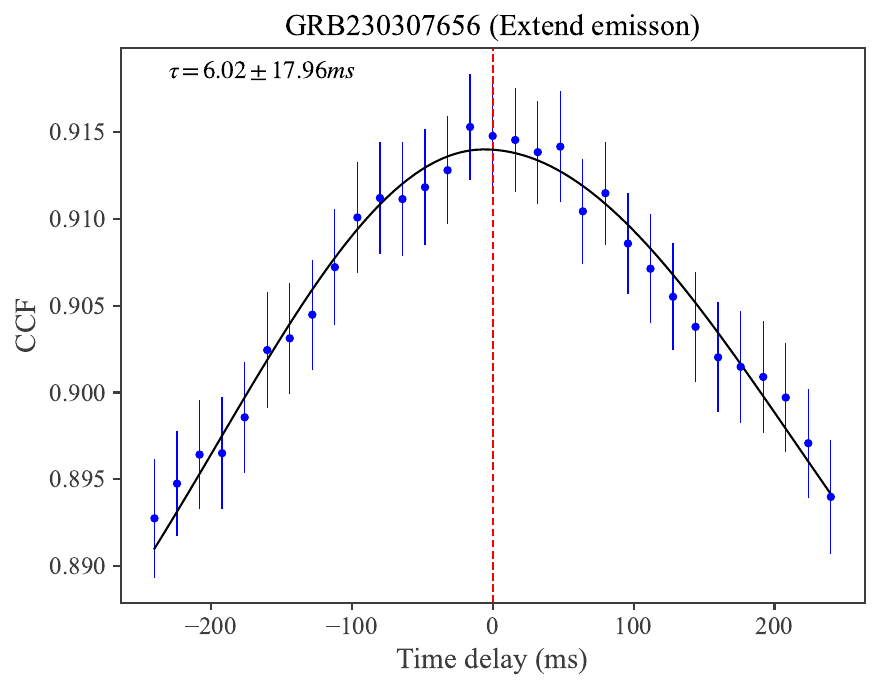}
\includegraphics[width=0.24\textwidth]{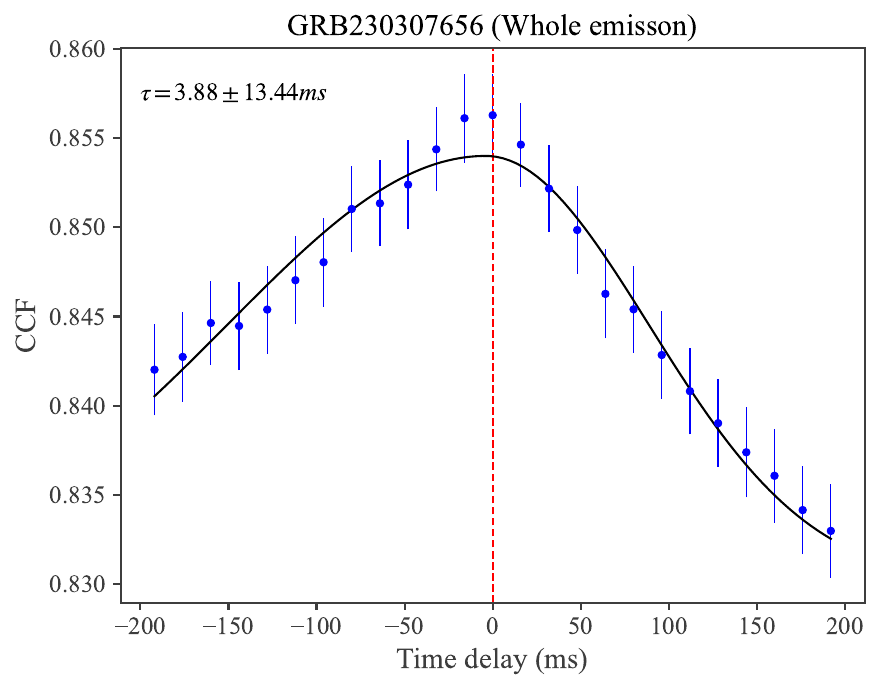}
\includegraphics[width=0.24\textwidth]{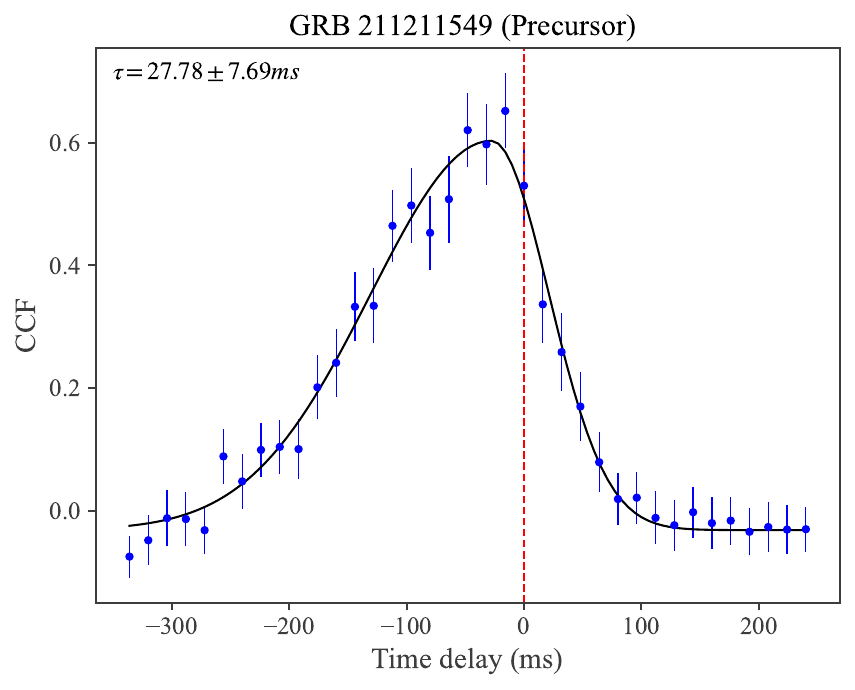}
\includegraphics[width=0.24\textwidth]{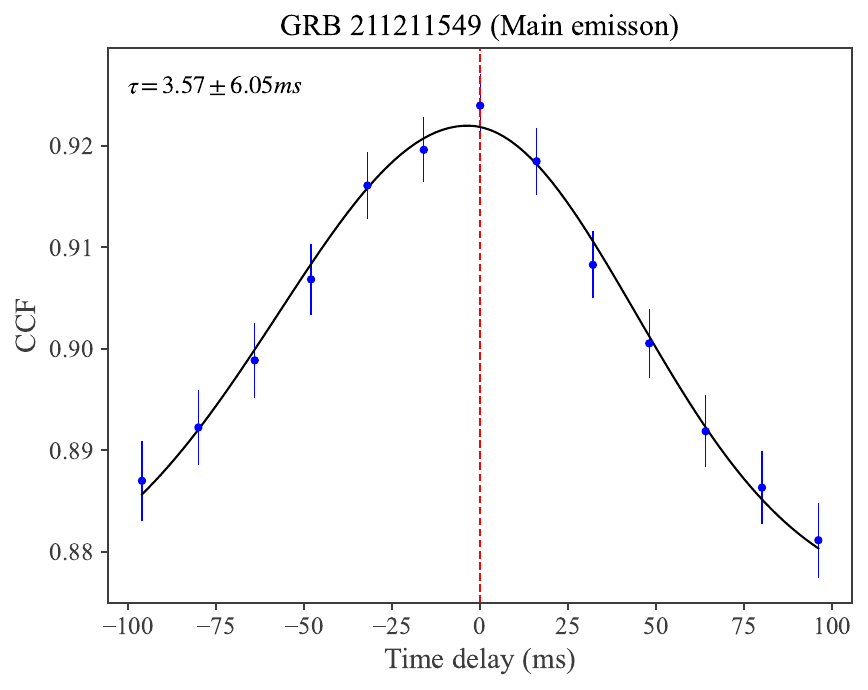}
\includegraphics[width=0.24\textwidth]{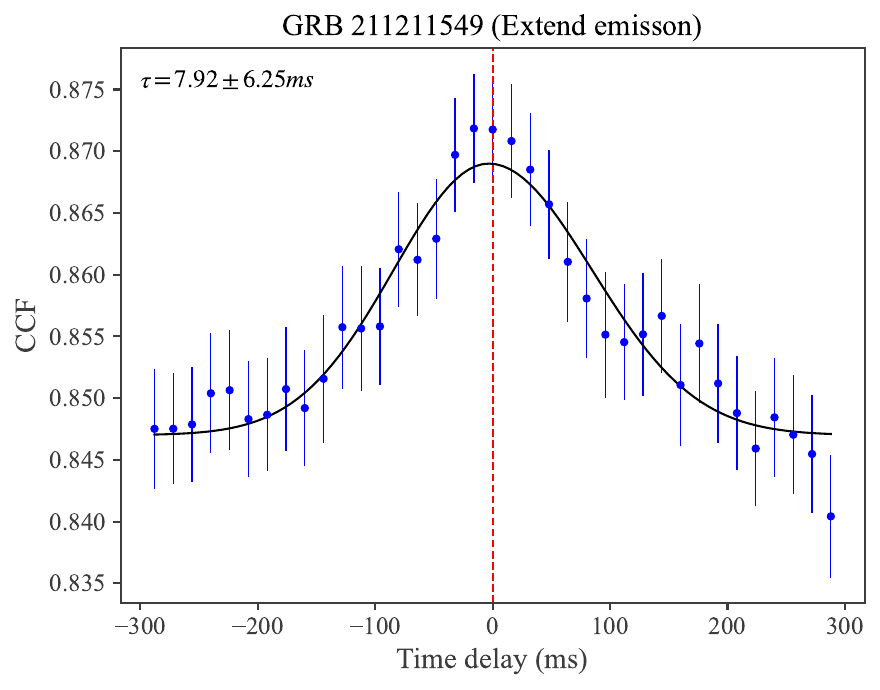}
\includegraphics[width=0.24\textwidth]{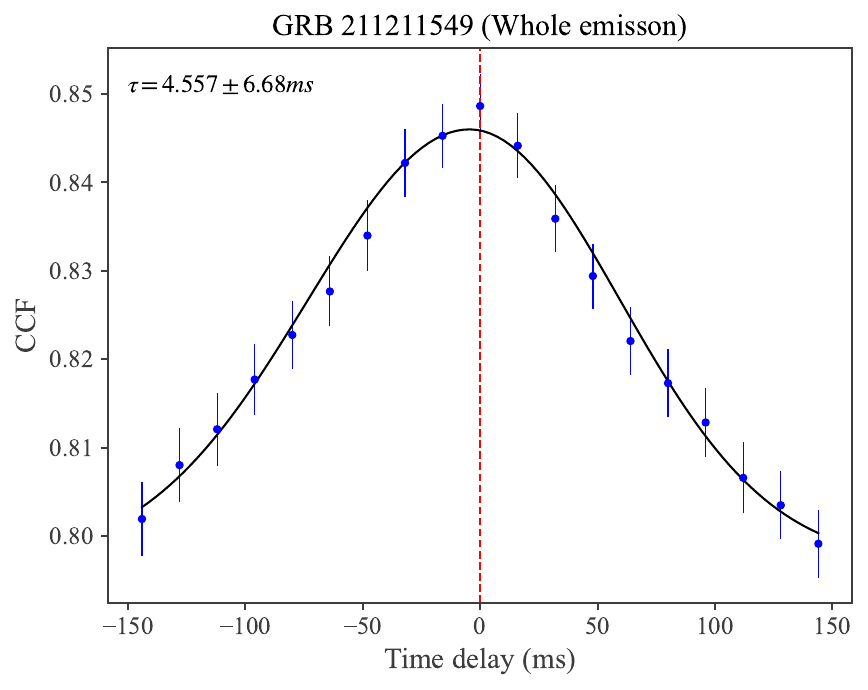}
\end{minipage}
\caption{Spectral lags of three episodes. All the time bin size of the three episodes of the two GRBs  is set to 0.008 s. The error comes from the Monte Carlo simulation of the light curve.  \label{fig2}}
\end{figure}

\subsubsection{The minimum variability timescale}
One intriguing feature in GRB temporal profile is the variability timescale, because the observed variability could be used to explore the variations of the central engine 
and the variability is imprinted on the radiation processes \citep{1997MNRAS.287..110S,2010ApJ...723..267M}. We calculate the minimum variability timescale (MVT) defined as the shortest duration of individual pulses in one profile. 
The MVT of the three phases of the two bursts are obtained by using the peak detection algorithm MEPSA to identify the shortest pulse within the two GRB lightcurves and estimate the full width at half maximum duration \citep{2023A&A...671A.112C}. The results are listed in Table 1. It is found that  the MVT of the two burtst are consistent in the full emission phase with MVT $\sim$ 39 ms and $\sim$ 49 ms, respectively.
The short MVTs also indicates the two GRBs fall into SGRB catalog since SGRBs have significantly shorter MVTs than long GRBs \citep{2023A&A...671A.112C}. Our result is consistent with that of others (e.g. \cite{2023A&A...671A.112C,2023yCat..36710112C}). We further compare the MVTs of the three phases in each burst. It is found that a smooth time profile is shown in precursors while the MVTs of the corresponding phase are also consistent (see, Table 1).
The $\rm{FWHM_{min}}$ of MVT vs $T_{90}$ of the two GRBs is shown in Figure 3. We can find that the two bursts lie in the same region and are consistent with the short-duration bursts with extended emission.

\begin{figure}[htbp]
\centering
\begin{minipage}[t]{0.6\textwidth}
\centering
\includegraphics[width=\textwidth]{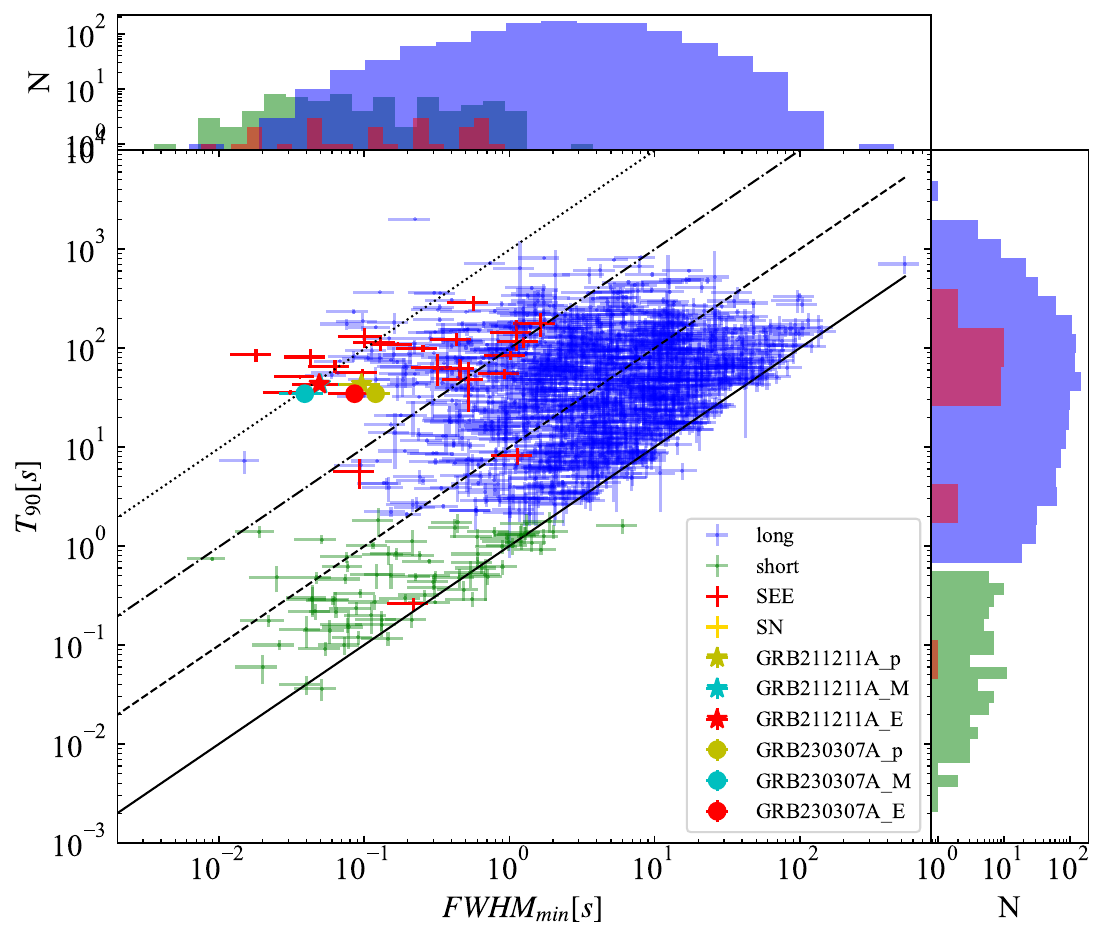}
\end{minipage}
\caption{The relationship between MVT and $T_{90}$ for GRB 230307A and GRB 211211A along with the short and long GRBs sample studied by \cite{2015ApJ...811...93G}.\label{fig3}}
\end{figure}

\begin{center}
\begin{longtable}{ccccccccccc}
\caption{A temporal and spectral properties comparison between GRB 230307A and GRB 211211A for the three stages}
\renewcommand\tabcolsep{0.9pt}
\renewcommand\arraystretch{0.8}
\label{table1}\\
stage& GRBname& Duration&Qt &MVT &lag & $\alpha$ &$\beta$ & $E_{p}$  & $E_{iso}$ \\
        &                 &     s         &  s   & ms          &  ms  &              &              &      keV      &  $10^{52}$ erg\\
\hline
\hline
\multirow{2}{*}{Precursor}&GRB 230307A& $\sim$ 0.30  & $\sim$ 0.30&120$\pm$ 40  & 24.11$\pm$2.49&${-0.59}^{+0.08}_{-0.01}$& ${-3.06}^{+0.21}_{-0.02}$&${168.85}^{+2.25}_{-10.42}$&1.20$\pm$0.12\\
&GRB 211211A & $\sim$ 0.20&$\sim$ 0.90 & 96 $\pm$30&27.78$\pm$7.69& ${-0.83}^{+0.14}_{-0.25}$& ${-2.75}^{+0.24}_{-0.16}$& ${46.1}^{+3.80}_{-3.62}$&0.06$\pm$0.01\\
\hline
\multirow{2}{*}{Main}&GRB 230307A& $\sim$ 17.40  & $\sim$2.00&39$\pm$ 13  & 4.73$\pm$11.43&${-0.82}^{+0.002}_{-0.003}$& ${-4.94}^{+0.03}_{-0.12}$&${897.33}^{+3.81}_{-2.53}$&150.10$\pm$3.47\\
&GRB 211211A & $\sim$ 13.00&$\sim$ 0.85 & 50 $\pm$20&3.57$\pm$6.05& ${-1.01}^{+0.002}_{-0.001}$& ${-2.39}^{+0.02}_{-0.02}$& ${739.47}^{+12.98}_{-10.69}$&49.94$\pm$0.14\\
\hline
\multirow{2}{*}{Extended}&GRB 230307A& $\sim$ 21.30  &  &86 $\pm$ 30  & 6.02$\pm$17.96&${-1.04}^{+0.04}_{-0.05}$& ${-3.72}^{+0.22}_{-0.13}$&${454.06}^{+5.21}_{-5.73}$&49.30$\pm$1.13\\
&GRB 211211A & $\sim$ 55.00& & 49 $\pm$20&7.92$\pm$6.25& ${-1.01}^{+0.002}_{-0.001}$& ${-2.03}^{+0.02}_{-0.01}$ & ${87.98}^{+4.76}_{-4.90}$&21.79$\pm$0.26\\
\hline
\multirow{2}{*}{Whole}&GRB 230307A& $\sim$ 2.00 &  &39 $\pm$ 13 & 3.88$\pm$13.44&${-1.05}^{+0.001}_{-0.002}$& ${-4.99}^{+0.01}_{-0.16}$&${844.72}^{+4.04}_{-3.15}$&195.28$\pm$5.64\\
&GRB 211211A & $\sim$ 1.00& & 49 $\pm$20&4.56$\pm$6.68& ${-1.28}^{+0.04}_{-0.05}$& ${-2.33}^{+0.05}_{-0.03}$ & ${87.98}^{+4.76}_{-4.90}$&71.89$\pm$1.00\\
\hline
\hline
\end{longtable}
\end{center}

\subsection{The spectral properties}
\subsubsection{The case of time-integrated spectral properties}
We perform time-integrated and detailed time-resolved spectral analysis for the two GRBs. Two empirical functions, Band and CPL introduced in the previous section, have been used to fit the time-integrated spectra. Through model comparison we can pick out the best model with the criterion $\Delta DIC$ = 10. This selection method was adopted by \citet{2022ApJ...932...25C}, \citet{2021ApJ...922...34C}, and \citet{2021ApJ...920...53C}. For GRB 211211A, our detailed spectral analysis shows that the time-integrated spectrum of the main burst and the extended emission can be fitted by a Band model, but the precursor can be fitted by a CPL model. On the contrary, for the GRB 230307A, the precursor can be modelled by the Band function, and both the main burst and the extended emission are fitted by a CPL model.

\begin{figure}[htbp]
\centering
\begin{minipage}[t]{0.9\textwidth}
\centering
\includegraphics[width=\textwidth]{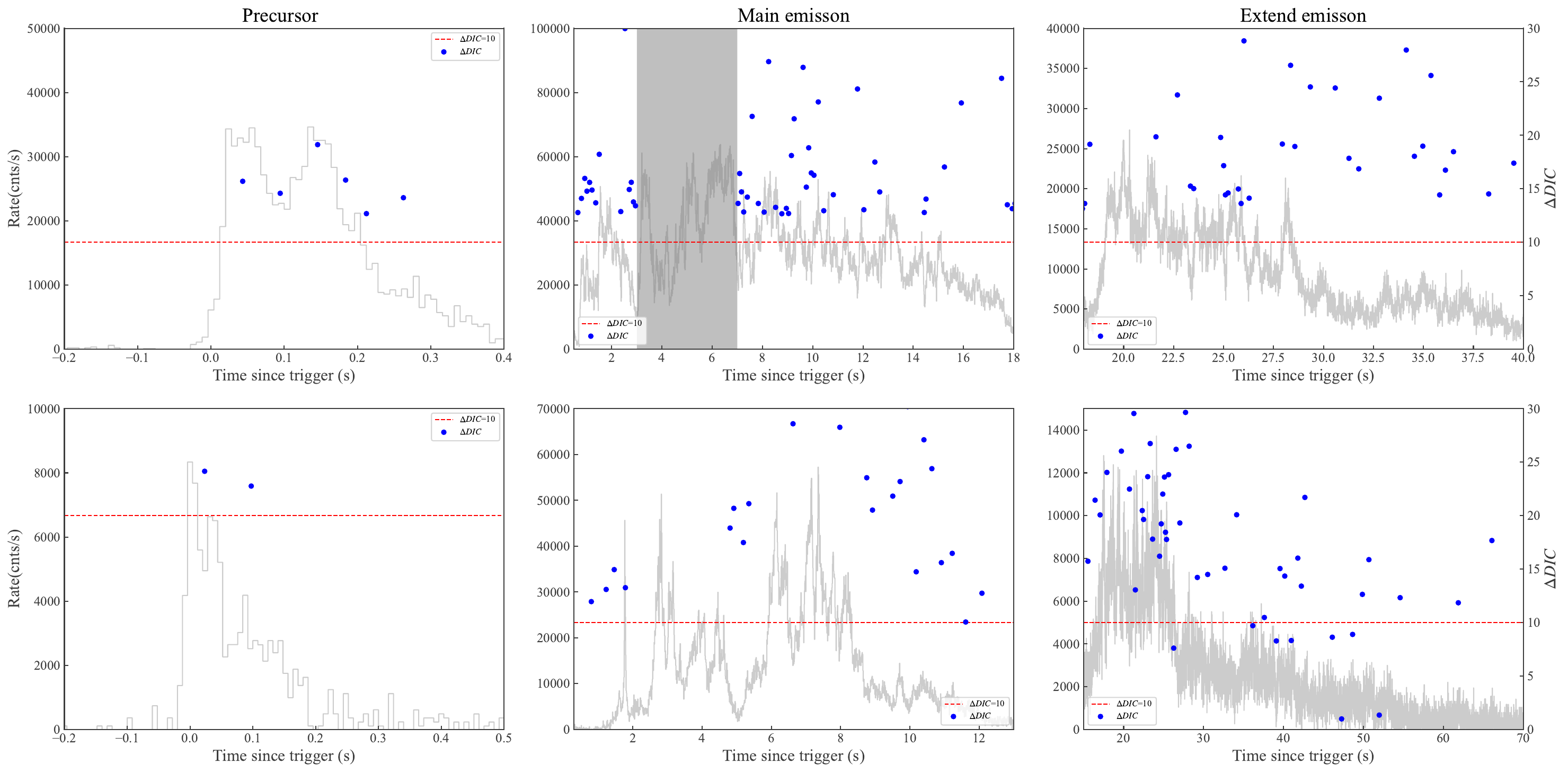}
\end{minipage}
\caption{Evolution of $\Delta DIC$ between the empirical model and empirical + BB models for GRB 230307A (upper three panels) and GRB 211211A (lower three panels). The grey curve represents the light curve, and the red dotted line represents $\Delta DIC$ = 10. If a DIC value lies above the red dotted line, then the presence of a BB component is favored. The grey shaded region in the upper second panel indicates the time interval of TTE data losses.\label{fig4}}
\end{figure}

\subsubsection{The case of the time-resolved spectral properties}
The spectral analysis of GRB 211211A provided by \cite{2023ApJ...943..146C} shows that the superposition models including non-thermal components and BB components are favored. Here, we first use two empirical functions, Band and CPL, to perform the time-resolved spectral analysis, respectively. We then add a BB component to the nonthermal model (Band+BB) or (CPL+BB) and compare the models by the DIC to check if there is any thermal component in the spectra. That is, $DIC_{CPL}$-$DIC_{CPL+BB}$ $>$ 10 or $DIC_{Band}$-$DIC_{Band+BB}$ $>$ 10 indicates that there is a thermal component. 

By performing the detailed spectral analysis of the two bursts, we obtain 138 and 135 time-resolved spectra for GRB 230307A and GRB 211211A, respectively. The fitting parameters of Band, Band+BB, CPL, and CPL+BB are given in online table. It is very interesting that almost all of the time-resolved spectra are quasi-thermal spectra. The exception is that the non-thermal feature is obtained from a few bins of the extended emission of GRB 211211A (see Figure 4 in detail). We further try to fit the spectra using single BB model and mBB model, respectively. However, the two models fail to fit the time-resolved spectra compared to the CPL+BB or the Band+BB model. Therefore, we identify that both thermal and non-thermal origin can co-exist in the two GRBs.

\begin{figure}[htbp]
\centering
\begin{minipage}[t]{0.9\textwidth}
\centering
\includegraphics[width=\textwidth]{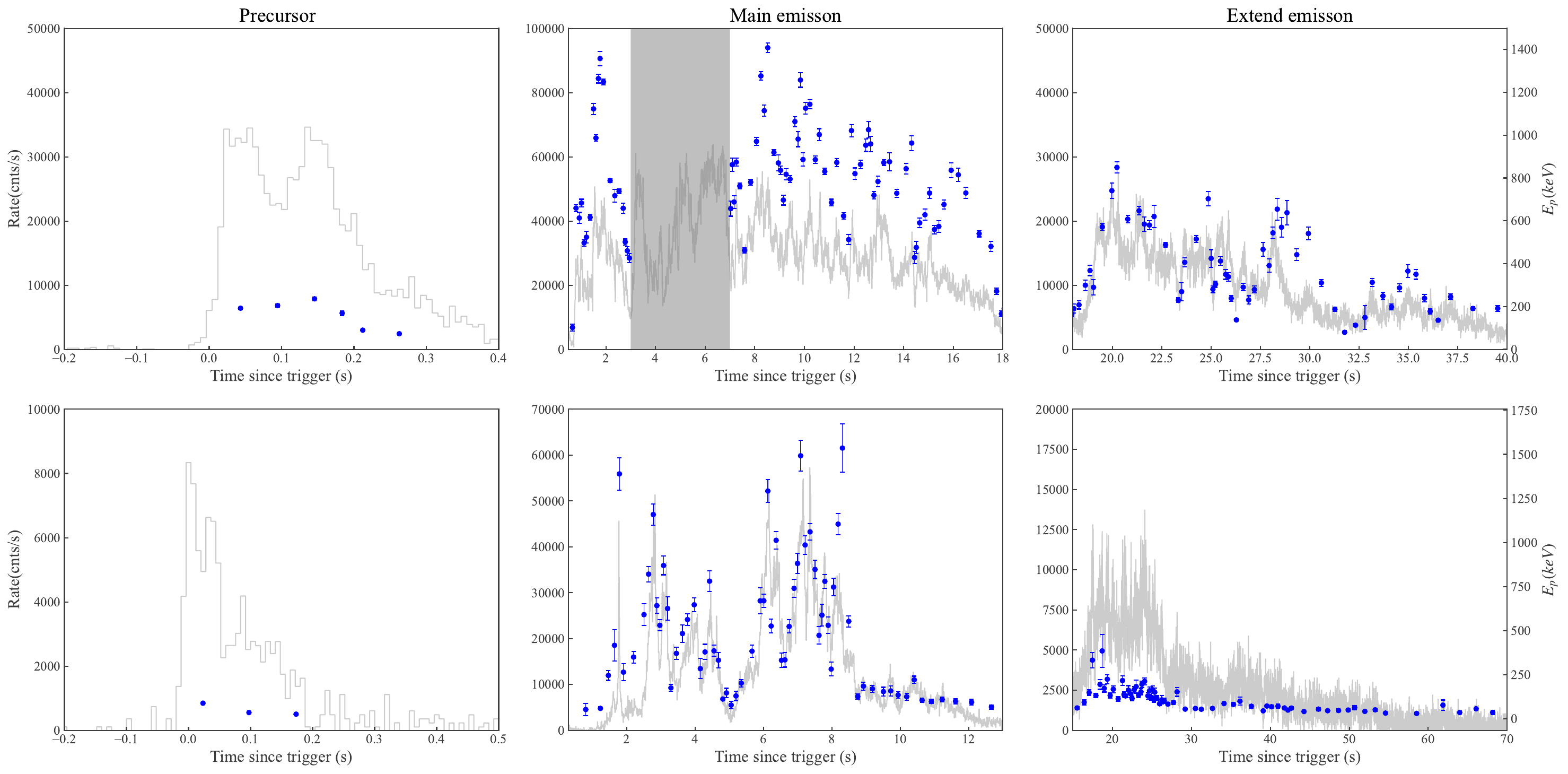}
\end{minipage}
\caption{The $E_{p}$ evolution of  three emission phases with time for the GRB 230307A (upper three panels) and GRB 211211A (lower three panels). The grey lines are the net light curve and the grey shaded region in the upper second panel indicates the time interval of TTE data losses. \label{fig5}}
\end{figure}

\begin{figure}[htbp]
\centering
\begin{minipage}[t]{0.9\textwidth}
\centering
\includegraphics[width=\textwidth]{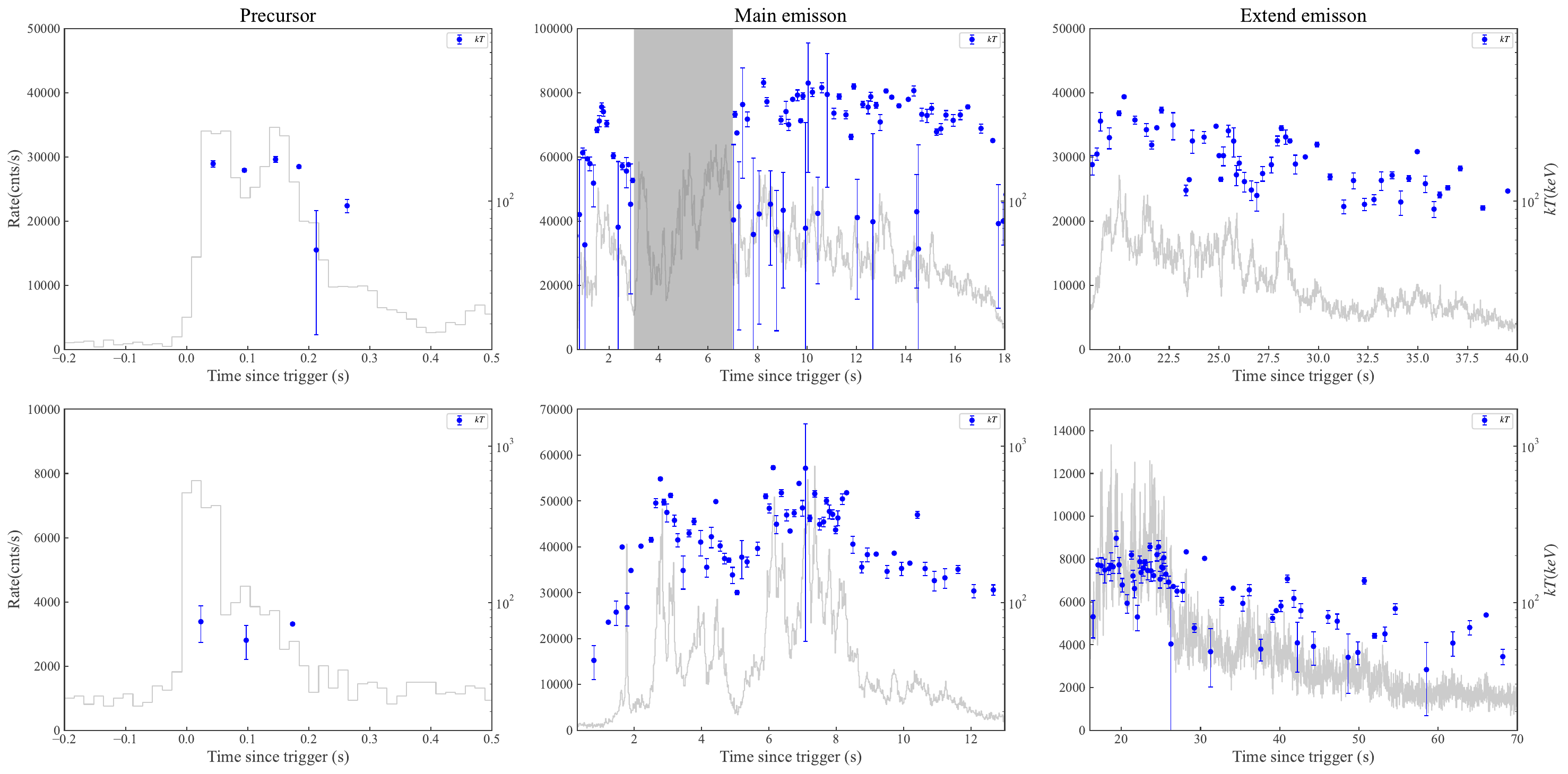}
\end{minipage}
\caption{The $kt$ evolution of  three emission phases with time for the GRB 230307A (upper three panels) and GRB 211211A (lower three panels). The grey lines are the net light curve and the grey shaded region in the upper second panel indicates the time interval of TTE data losses. We find that the best model CPL+BB can not well constrain all the blackbody spectra for GRB 230307A,  so as we see some kT values for GRB 230307A having very large errors.\label{fig6}}
\end{figure}

\begin{figure}[htbp]
\centering
\begin{minipage}[t]{0.9\textwidth}
\centering
\includegraphics[width=\textwidth]{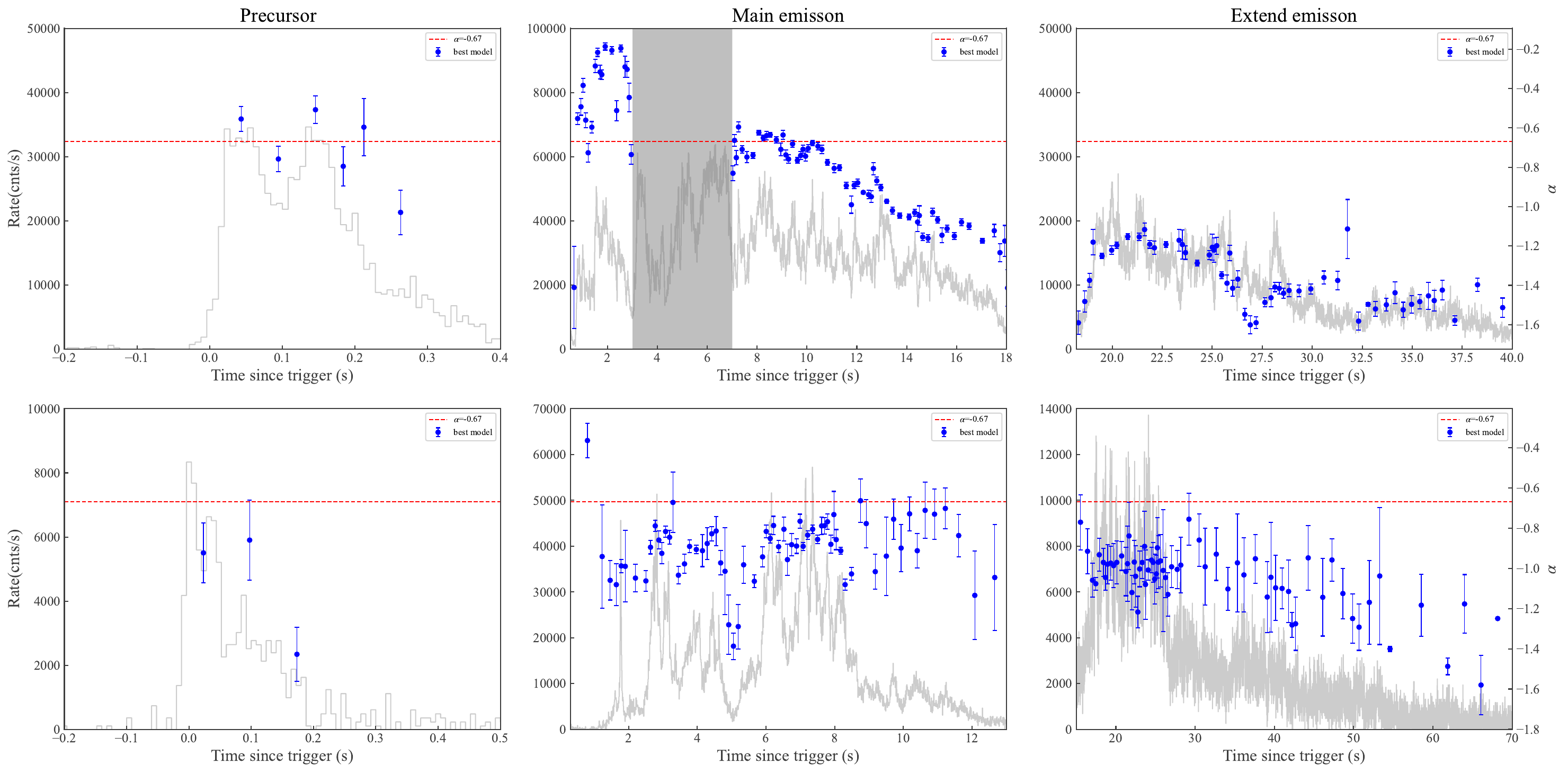}
\end{minipage}
\caption{The $\alpha$ evolution of  three emission phases with time for the GRB 230307A (upper three panels) and GRB 211211A (lower three panels). The grey lines are the net light curve and the grey shaded region in the upper second panel indicates the time interval of TTE data losses.  \label{fig6}}
\end{figure}

\begin{figure}[htbp]
\centering
\begin{minipage}[t]{0.9\textwidth}
\centering
\includegraphics[width=\textwidth]{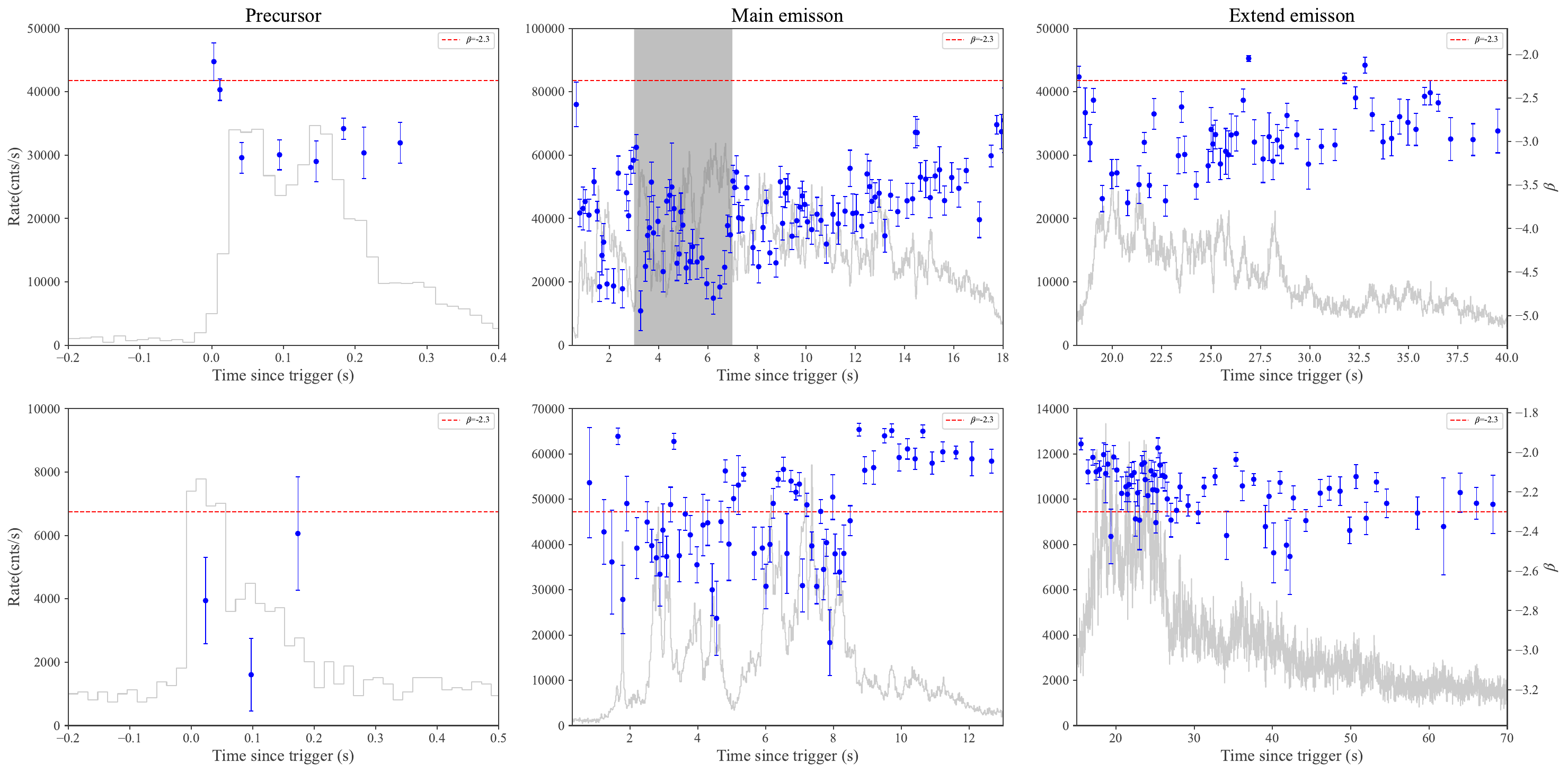}
\end{minipage}
\caption{The $\beta$ evolution of  three emission phases with time for the GRB 230307A (upper three panels) and GRB 211211A (lower three panels). The grey lines are the net light curve and the grey shaded region in the upper second panel indicates the time interval of TTE data losses.  \label{fig6}}
\end{figure}

We can investigate how the modelling parameters evolve with time by looking at the time-resolved spectra. The results are given in Figures 5, 6, 7, and 8. It is found that both the evolution of $E_{p}$ and $\alpha$ show flux tracking mode, that is, the parameter $E_{p}$ and $\alpha$ are related to the rise and the fall of flux. This is also called “Double-tracking” \citep{2019ApJ...884..109L}. It is very interesting that GRB 230307A and GRB 211211A share similar “Double-tracking” modes. The kt also tracks the flux in Figure 6. However, the $\beta$ does not show evident evolution with time (see Figure 8). In Figure 7 we also find that the numbers of $\alpha$ obtained by the Band model and the CPL model in the time-resolved spectrum exceed the so-called ``death line" \citep{2002ApJ...581.1248P}.

We further investigate the relations between $E_{p}$ and flux for the two GRBs exhibits linear correlations in logarithmic space for the three phases. Moreover, the whole emission for the two GRBs shows a very similar correlation and slope (see Figure 9 and Table 2). Similar correlations between $\alpha$ and flux as well as $\alpha$ and $E_{p}$ also exist but the correlations are weaker than that of $E_{p}$ and flux (see Figures 10 and 11 and Table 2). The weaker correlation might be explained by the fact that $\alpha$-F correlation exhibit an obvious broken behaviour for GRB 230307A \citep{2023arXiv230311083W}. The correlations between $E_{p}$ and F as well as $\alpha$ and F support the “Double-tracking” mode. 

\begin{figure}[htbp]
\centering
\begin{minipage}[t]{0.9\textwidth}
\centering
\includegraphics[width=\textwidth]{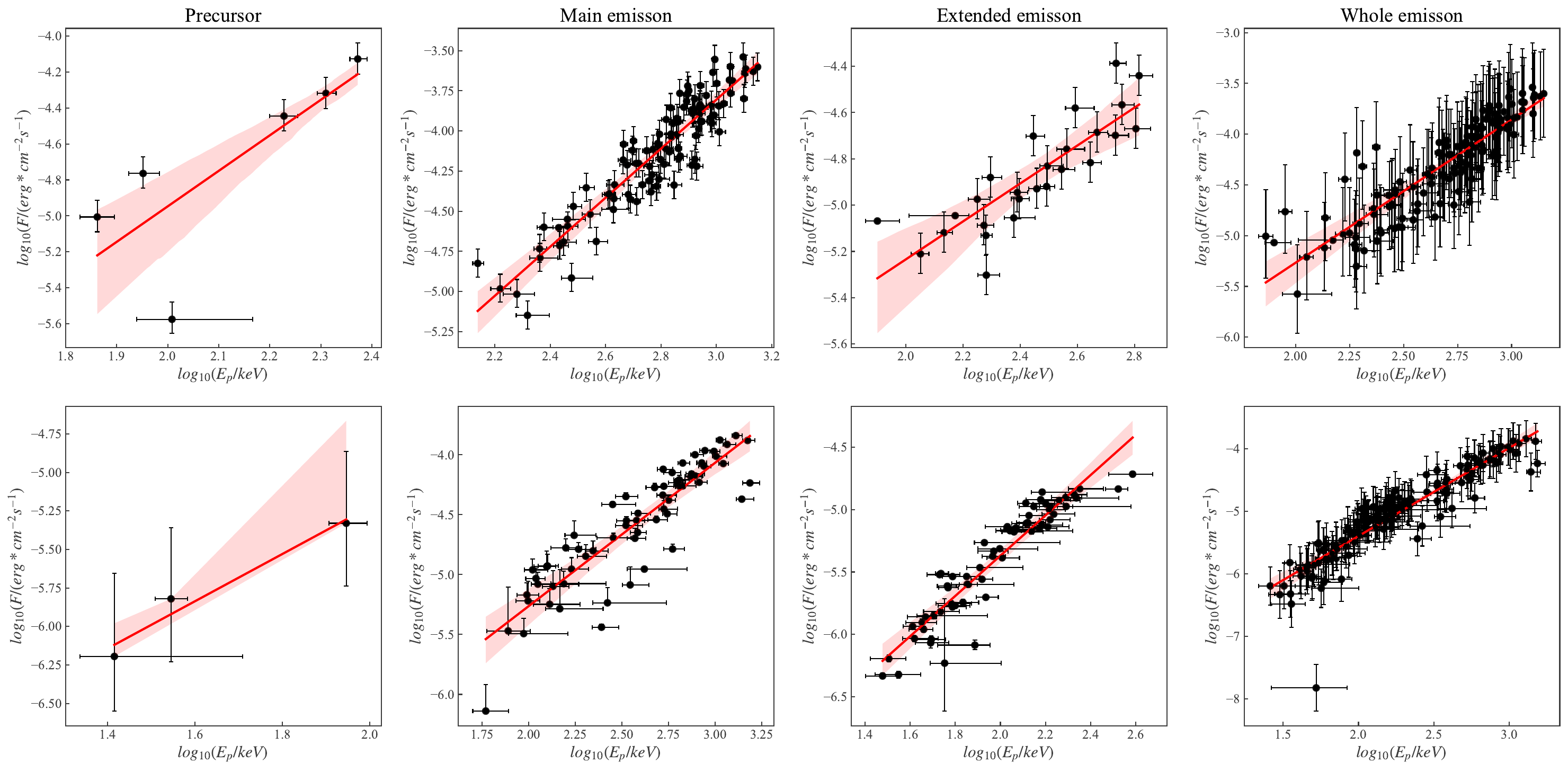}
\end{minipage}
\caption{The $E_{p}$ - $F$ relation for GRB 230307A and GRB 211211A. The upper four panels represent the precursor, main emission, extended emission and the whole emission of GRB 230307A, while lower four panels are the corresponding emission phase for GRB 211211A.  \label{fig7}}
\end{figure}

\begin{figure}[htbp]
\centering
\begin{minipage}[t]{0.9\textwidth}
\centering
\includegraphics[width=\textwidth]{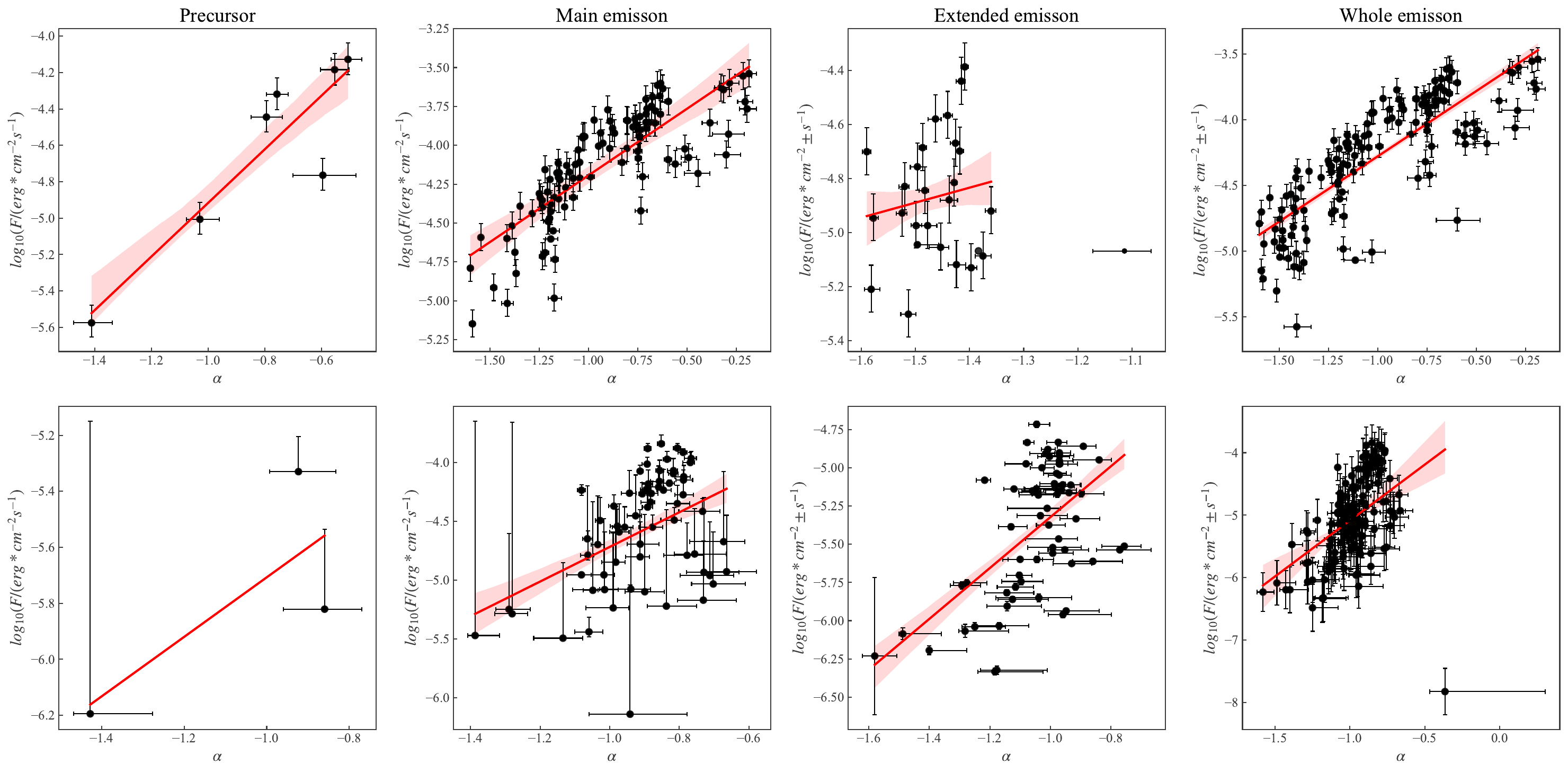}
\end{minipage}
\caption{The $\alpha$ - $F$ relation for GRB 230307A and GRB 211211A. The upper four panels represent the precursor, main emission, extended emission and the whole emission of GRB 230307A, while lower four panels are the corresponding emission phase for GRB 211211A.  \label{fig8}}
\end{figure}

\begin{figure}[htbp]
\centering
\begin{minipage}[t]{0.9\textwidth}
\centering
\includegraphics[width=\textwidth]{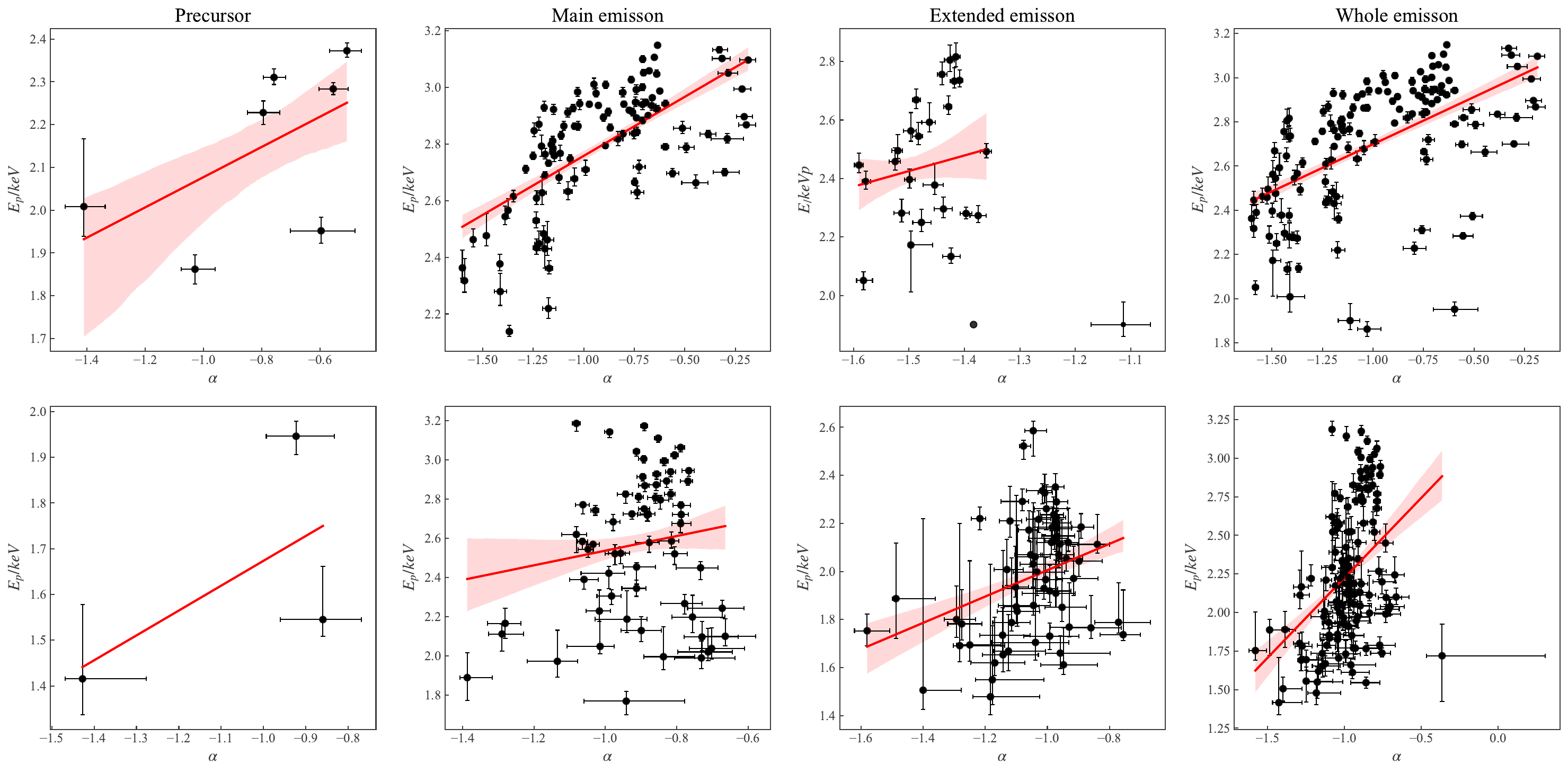}
\end{minipage}
\caption{The $\alpha$-$E_{p}$ relation for GRB 230307A and GRB 211211A. The upper four panels represent the precursor, main emission, extended emission and the whole emission of GRB 230307A, while lower four panels are the corresponding emission phase for GRB 211211A.  \label{fig9}}
\end{figure}

\begin{deluxetable}{cccccccccccccc}
\tablecaption{The correlation analysis results of spectral parameters.\label{}}
\tablehead{
\multirow{2}{*}{Relation} & \multirow{2}{*}{GRB}& &\multicolumn{2}{c}{Precursor}& &\multicolumn{2}{c}{Main}& &\multicolumn{2}{c}{Extended}& &\multicolumn{2}{c}{Whole}\\
    \cline{4-5}\cline{7-8}\cline{10-11} \cline{13-14}
    & & &$\rho$ & a& &$\rho$ &a & &$\rho$&a& &$\rho$&a\\
}
\startdata   
\hline
\multirow{2}{*}{$log_{10}E_p - log_{10}F$}&GRB 230307A& &0.83&$1.99\pm0.04$& &0.90&$1.61\pm0.10$& &0.91&$1.30\pm0.07$& &0.91&$1.46\pm0.03$\\
&GRB 211211A& &1.00&$1.93\pm0.09$& &0.90&$1.28\pm0.03$& &0.95&$1.69\pm0.04$& &0.96&$1.43\pm0.01$\\
\hline
\multirow{2}{*}{$\alpha - log_{10}F$}&GRB 230307A& &0.89&$1.49\pm0.08$& &0.84&$0.91\pm0.02$& &0.10&$0.56\pm0.03$& &0.83&$1.02\pm0.02$\\
&GRB 211211A& &0.50&$1.42\pm0.07$& &0.39&$1.48\pm0.67$& &0.46&$1.59\pm0.06$& &0.58&$1.73\pm0.04$\\
\hline
\multirow{2}{*}{$\alpha - log_{10}E_p$}&GRB 230307A& &0.64&$0.35\pm0.07$& &0.65&$0.50\pm0.03$& &0.13&$0.55\pm0.09$& &0.62&$0.48\pm0.02$\\
&GRB 211211A& &0.50&$0.54\pm0.03$& &0.07&$0.37\pm0.08$& &0.28&$0.54\pm0.06$& &0.42&$1.08\pm0.04$\\
\hline
\multirow{2}{*}{$log_{10}F_{BB} - log_{10}F$}&GRB 230307A& &0.90&$1.50\pm0.10$& &0.82&$0.47\pm0.03$& &0.70&$0.46\pm0.01$& &0.87&$0.63\pm0.10$\\
&GRB 211211A& &0.50&$0.67\pm0.08$& &0.82&$0.70\pm0.11$& &0.91&$0.54\pm0.02$& &0.93&$0.69\pm0.02$\\
\hline
\enddata
\tablecomments{The $\rho$ and a are the Spearman rank correlation Coefficient and the slope of the best fit line.}
\end{deluxetable}

\subsection{Amati relation\label{3.1}}
The well-known Amati relation \citep{2002A&A...390...81A} is a spectrum-energy correlation, and it can be used for diagnosing GRB classification (e.g. \cite{2013MNRAS.430..163Q}) as typical long and short GRBs follow two separate trajectories in the correlation plot. We compare the Amati relation for the three-stage emissions of the two bursts. We perform time-integrated spectral analysis of the three phases for the two bursts adopting the Band function. Our results, shown in Figure 12, prove that (1) both the whole emission of the two bursts are close to the Type II region, (2) both the precursors of the two bursts follow the Type-I rather than Type-II GRB track, (3) For the case of main emission, the two GRBs seem to locate in the transition region, (4) the extended emission show two different trends,  GRB 211211A follows the Type-II GRB track, while GRB 230307A is in the transition region.   

\begin{figure}[htbp]
\centering
\begin{minipage}[t]{\textwidth}
\centering
\includegraphics[width=0.65\textwidth]{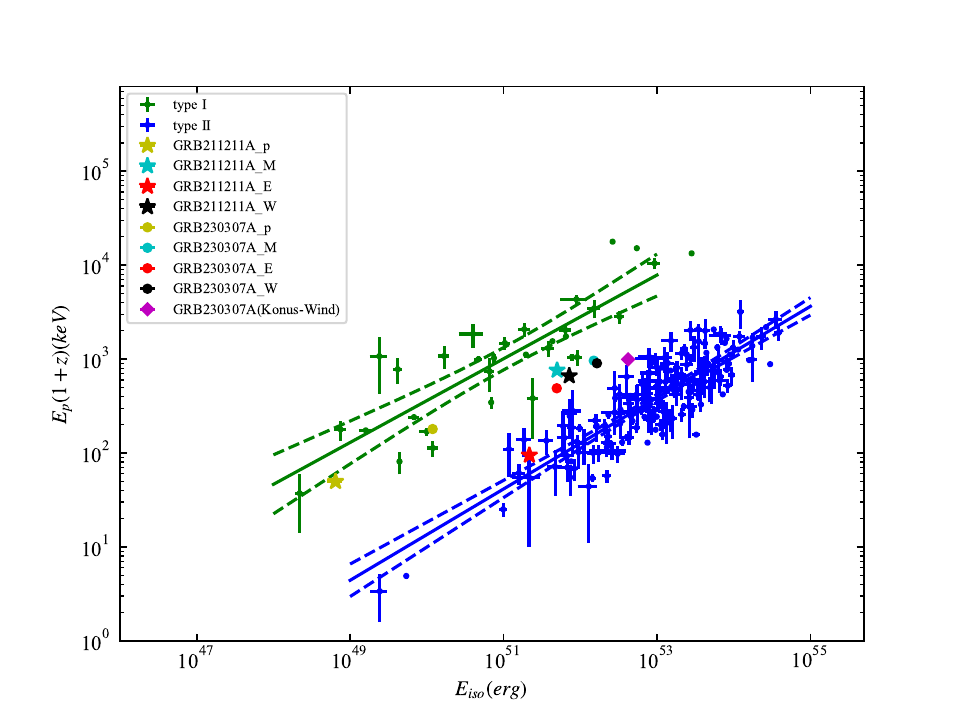}
\end{minipage}
\caption{Amati relationship diagram. Blue and red dashed lines represent the best-fit correlations for the type-I and type-II GRB samples, respectively.  \label{fig10}}
\end{figure}

\subsection{hardness ratio relation}
Some studies show that hardness ratio (HR) defined by the fluence (flux) ratio in harder to softer energy bands can be used to classify GRBs (e.g. \cite{2006A&A...447...23H,2023MNRAS.518.6243J}). Short GRBs are harder with larger values of HR compared to the long ones. HR was found to be correlated with $T_{90}$ in one GRB sample (e.g. \cite{2000PASJ...52..759Q}). Here, we estimate the HR using flux ratio in two different energy channels: 10-50 keV and 50-300 keV energy bands of Fermi-GBM. We first fit the sample mentioned above with the Gaussian mixture model (GMM) and obtain two types of bursts based on BIC  (Bayesian Information Criterion) values. Then we use a machine learning algorithm Bayesian Gaussian mixture model (BGM) supported by scikit-learn \citep{2011JMLR...12.2825P} to estimate the probability of the two GRBs. The probabilities of the two bursts as LGRB are 99.8\% and 99.9\%, respectively. The HR-duration plot of Fermi-GBM data along with GRB 230307A and GRB 211211A is shown in Figure 13. Although GRB 211211A and GRB 230307A show some short burst features as presented in Section 3.1.2, they tend to better agree with the long burst population in the Amati relation. In order to further investigate if the three phases are different, we also analyse the relation of the three stages separately. Using the same method as the whole bursts we find that the precursors for both the two bursts locate in the region of SGRBs and the whole emission including main and the extended emission are still in agreement with the LGRB distribution, which is very consistent with the case of the Amati relation.

\begin{figure}[htbp]
\centering
\begin{minipage}[t]{\textwidth}
\centering
\includegraphics[width=0.65\textwidth]{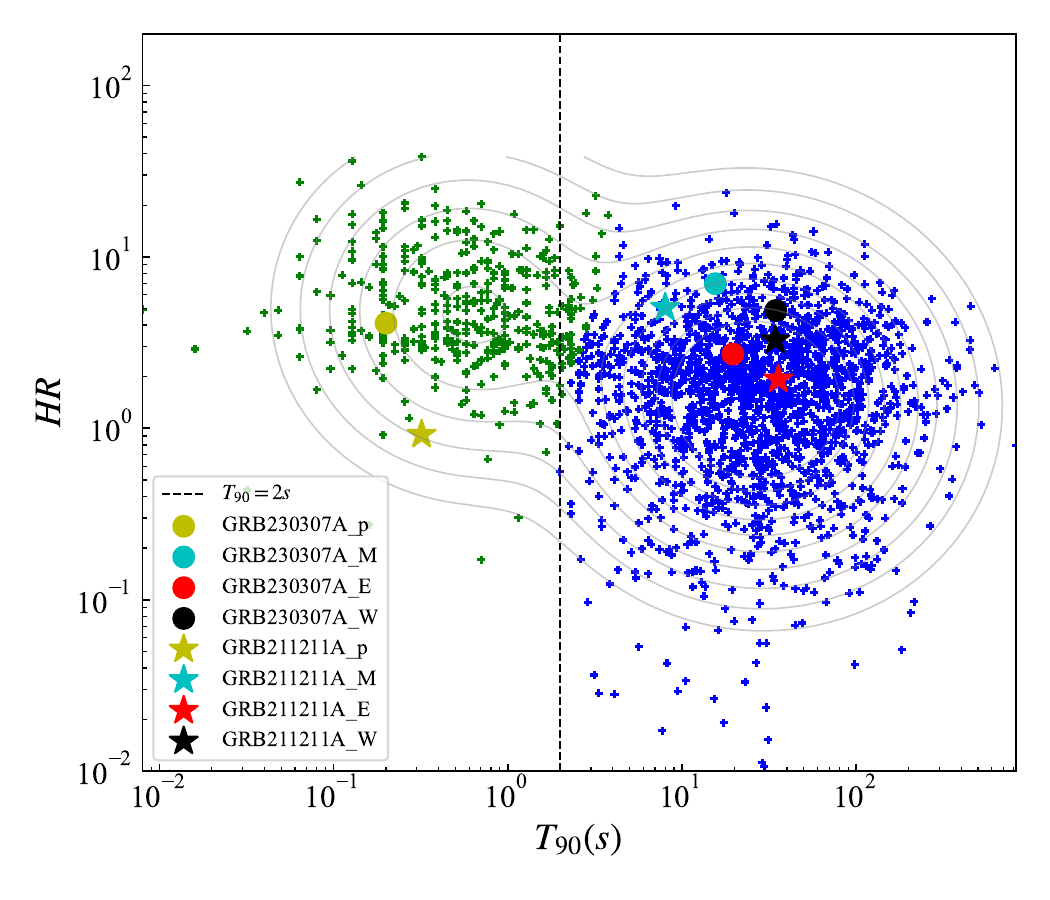}
\end{minipage}
\caption{The spectral hardness and $T_{90}$ for GRB230307A and GRB 211211A along with the GRB sample (including short and long GRBs) with the dataset available from Fermi-GBM catalog.  \label{fig11}}
\end{figure}

\subsection{Spectral Width}
The spectral width of the observed GRB spectrum has been used as a new tool to infer the physical properties of GRB emission by spectral fitting of empirical models \citep{2012ApJ...752..132P,2015MNRAS.447.3150A,2019A&A...629A..69B}. \cite{2015MNRAS.447.3150A} define the width as the log ratio of the energy at the full width at half maximum (FWHM) spectrum:
\begin{equation}
W={{\log }_{10}}\left( \frac{{{E}_{2}}}{{{E}_{1}}} \right)
\end{equation}
where $E_1$ and $E_2$ are the lower and upper energy limits of the $FWHM$ range, respectively. W depends only on the Band function parameters $\alpha$ and $\beta$, in order for the band function to have a peak  in the $\nu f_{\nu}$ representation, $\alpha$ must be greater than $-2$ and $\beta$ must be less than $-2$.

We calculate the width of each time-resolved spectrum based on the fitting results of the Band function. As shown in Figures 14 and 15, GRB 230307A has narrower spectra than GRB 211211A and the two GRBs have wider spectra than those of SGRB (the median number is $\sim0.86$ for Fermi-GBM short bursts \citep{2015MNRAS.447.3150A}). Therefore, the two bursts seem to belong to LGRBs based on the fact that the spectral width of long and short GRBs are significantly different \citep{2015MNRAS.447.3150A}. Moreover, the spectral widths of GRB 230307A and GRB 211221A gradually increase with time. \cite{2015MNRAS.447.3150A} computed the spectral widths by thermal radiation or synchrotron radiation. It was found that the thermal spectra are much narrower than synchrotron radiation spectra. Moreover, some evidences show that GRB composition changes from a fireball to a Poynting-flux-dominated jet based on the transition of the observed spectra from thermal to non-thermal emission (e.g. \citet{2018NatAs...2...69Z} and \citet{2022ApJ...940...48D}). Our results seem to support the GRB transition from fireball to Poynting flux based on the fact that the observed spectral widths  increase with time. 

\begin{figure}[htbp]
\centering
\begin{minipage}[t] {\textwidth}
\centering
\includegraphics[width=0.6\textwidth]{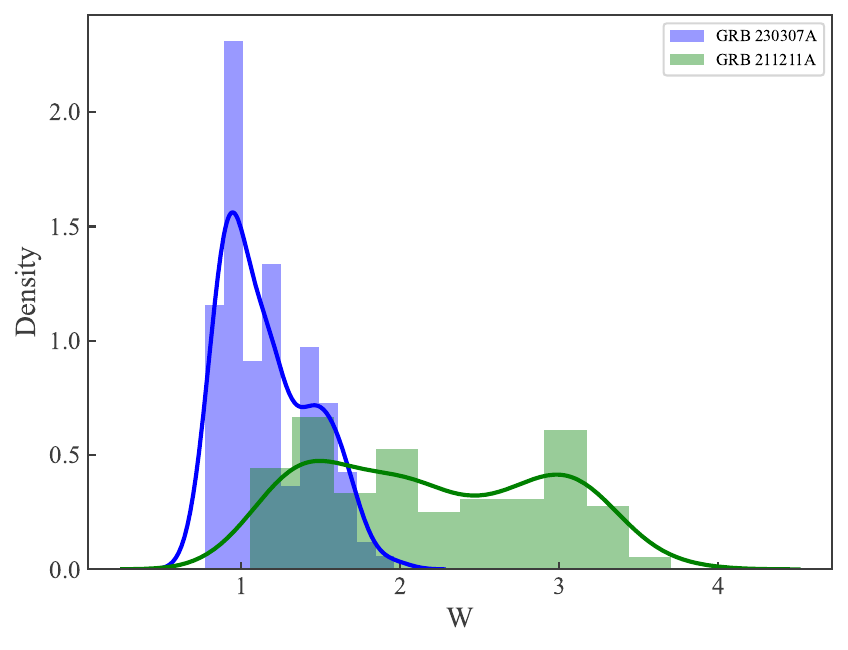}
\end{minipage}
\caption{The histogram of the spectral width for the two GRBs. Blue and green lines represent the kernel density for the GRB 230307A and GRB 211211A, respectively. The higher the peak in the kernel density plot, the more "denser" the data here, and the peak widths of the GRB 211211A are much larger than those of the GRB 230307A.  \label{fig12}}
\end{figure}

\begin{figure}[htbp]
\centering
\begin{minipage}[t] {\textwidth}
\centering
\includegraphics[width=\textwidth]{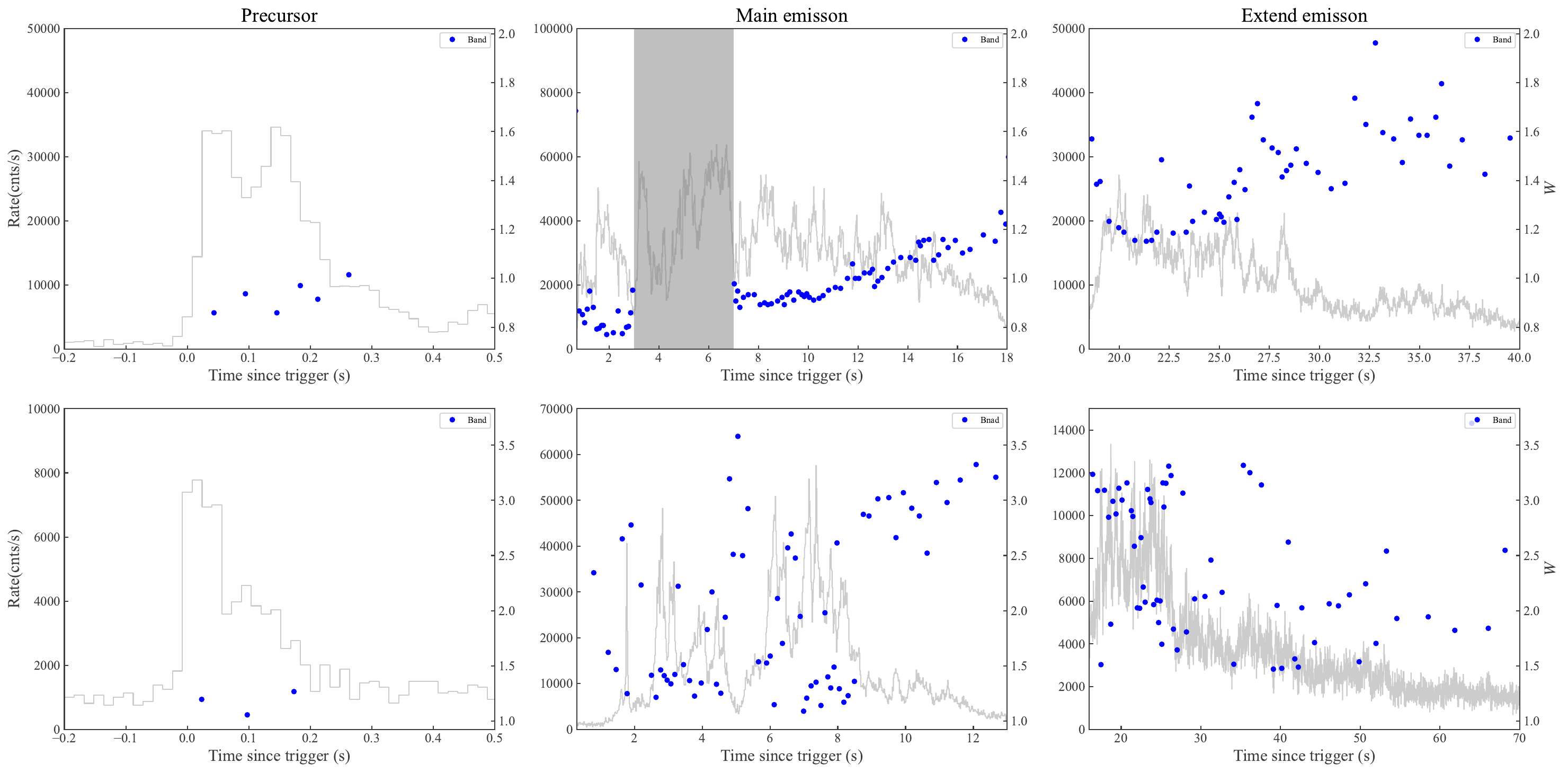}
\end{minipage}
\caption{The evolution of the spectral width with time for the three phases. The upper three panels represent the precursor, main emission, extended emission of GRB 230307A, while lower three panels are the corresponding emission phase for GRB 211211A.\label{fig13}}
\end{figure}

\subsection{The evolution of the thermal emission components}
In order to further check if the thermal emission components weaken with time we investigate if the thermal emission flux ($F_{bb}$) decreases with time since the thermal components exist through the emission phase. We find there are evident trends that $F_{bb}$ track the total flux and the overall trend is that the $F_{bb}$ decrease with time (see, Figure 16). That is, the thermal flux decreases with the decrease of the total flux rather than only the thermal flux decrease and the non-thermal flux increase. Figure 17 demonstrates the strong correlations between $F_{bb}$ and total flux $F$ and the Spearman-rank correlation coefficients as well as the slopes for the corresponding stages are very consistent for the two bursts (see, Table 2). These show that thermal emission components proportionally contribute the energy and the ratios of $F_{bb}$ to $F$ are unchanged through emission phase. Therefore, the fact show the two GRBs are indeed quasi-thermal and their jets do not transition from fireball to Poynting flux.

\begin{figure}[htbp]
\centering
\begin{minipage}[t]{0.9\textwidth}
\centering
\includegraphics[width=\textwidth]{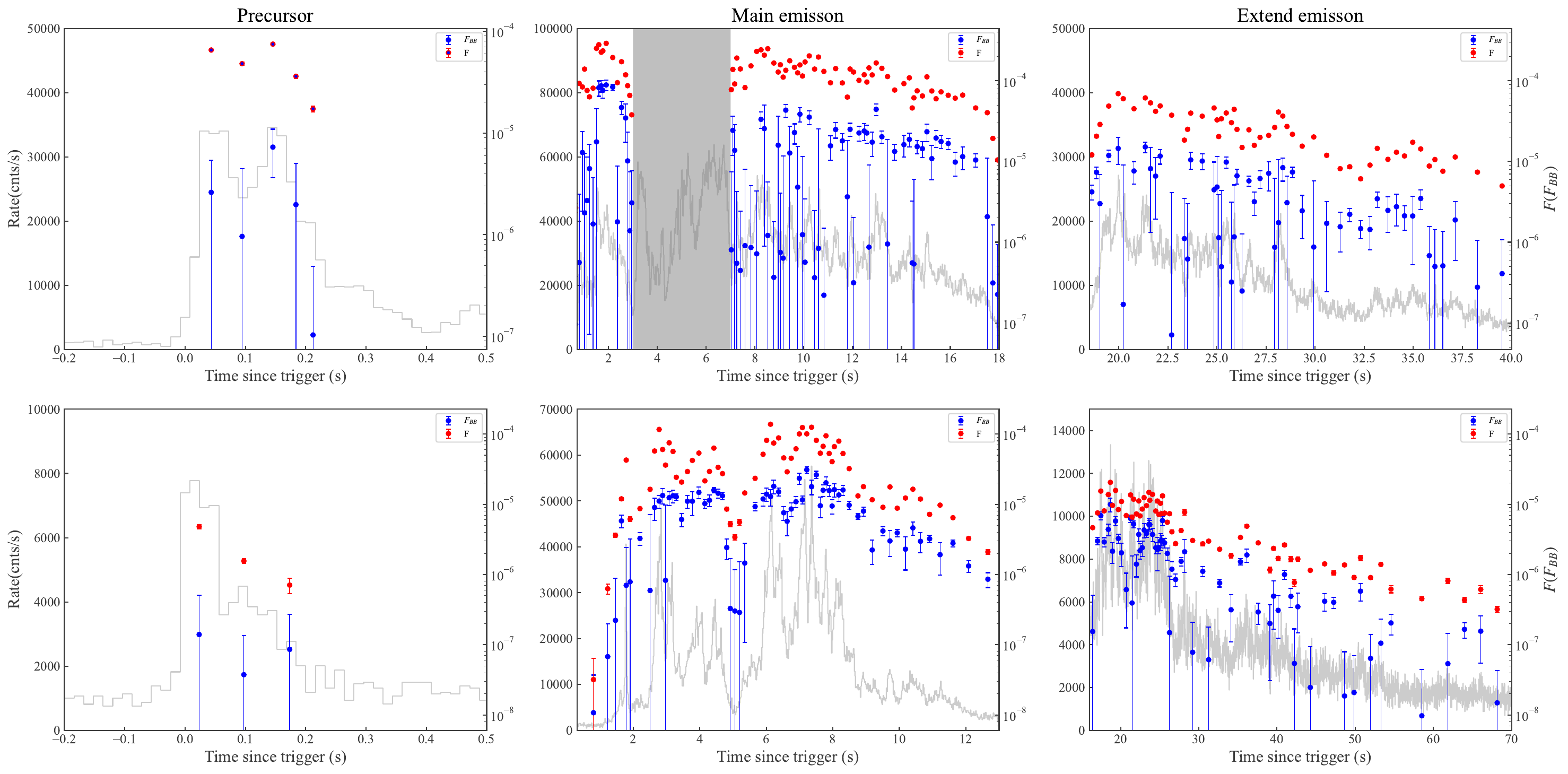}
\end{minipage}
\caption{The evolution of $F_{bb}$ with time for the GRB 230307A and GRB 211211A,where the red and the blue data points denote the total flux and the thermal flux, respectively. The upper three panels represent the precursor, main emission, extended emission of GRB 230307A, while lower three panels are the corresponding emission phase for GRB 211211A.  \label{fig14}}
\end{figure}

\begin{figure}[htbp]
\centering
\begin{minipage}[t]{0.9\textwidth}
\centering
\includegraphics[width=\textwidth]{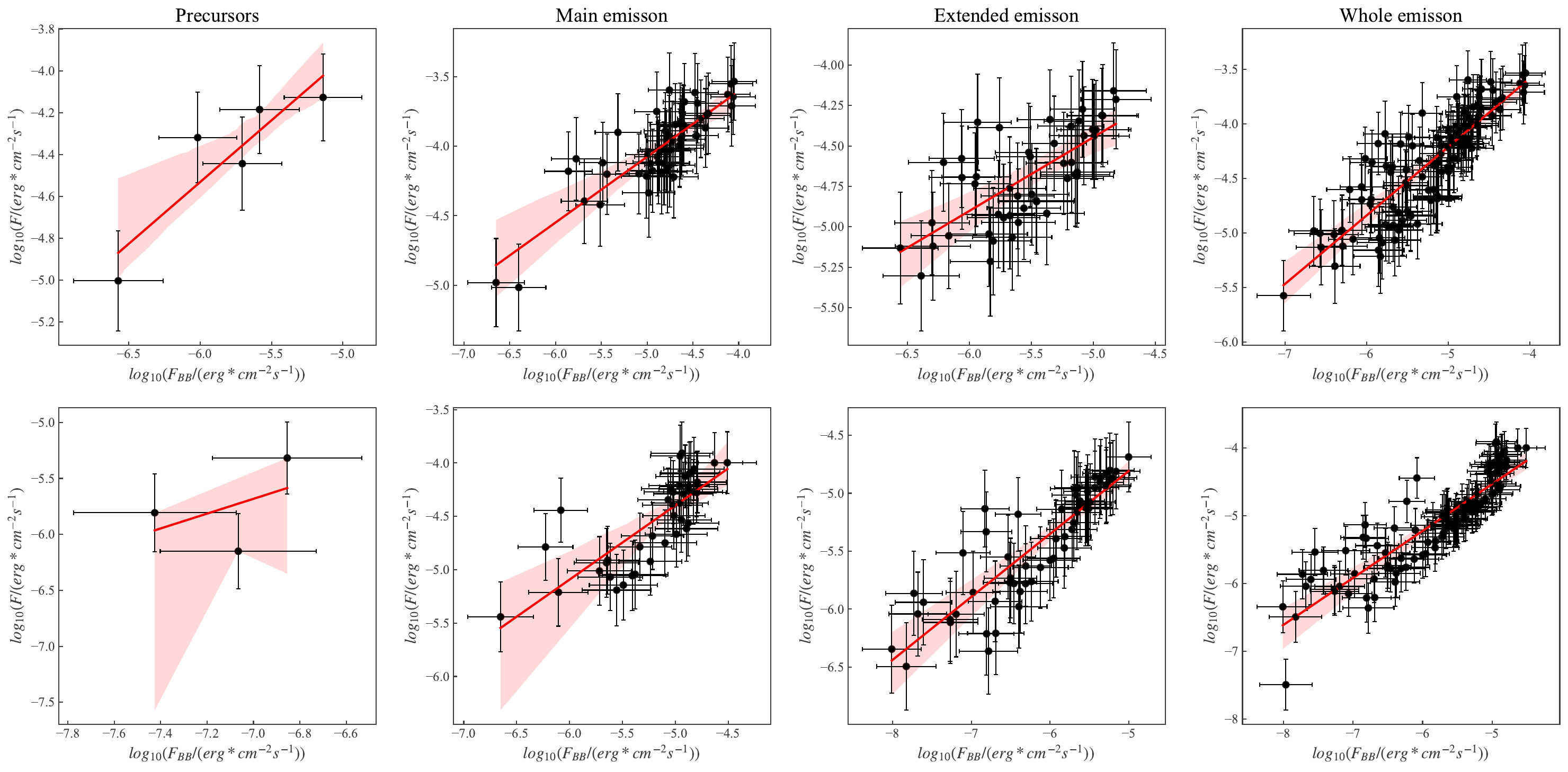}
\end{minipage}
\caption{The $F$-$F_{bb}$ relation for GRB 230307A and GRB 211211A. The upper four panels represent the precursor, main emission, extended emission and the whole emission of GRB 230307A, while lower four panels are the corresponding emission phase for GRB 211211A. \label{fig15}}
\end{figure}

\section{The photosphere emission properties\label{photosphere emission}}
The time-resolved spectra of the two bursts are quasi-thermal. The observational properties allow us to investigate the physical properties of the photoshperic radiation in relativistic outflow, including the bulk Lorentz factor $\Gamma$, the photospheric radius $r_{ph}$, the initial size of the outflow $r_0$, and the saturation radius $r_{s}$ \citep{2007ApJ...664L...1P}. Based on the photosphere emission model, when the fireball expands rapidly, the thermal energy of photons will be converted into the kinetic energy of the baryons from an initial radius $r_0$. According to the conservation of energy and entropy, the bulk Lorentz factor of the outflow increases with radius as  $\Gamma \propto r$ before the saturation radius $r_{s}=\Gamma \times r_0$ \citep{2019ApJS..242...16L}. Therefore, we can deduce the physical parameters once the thermal component is identified in the observed spectrum. With the known redshift information for the two bursts, some outflow parameters can be also estimated. 

The ratio between the observed thermal emission flux and the black body temperature (denoted as $\mathcal{R}$) under spherical symmetry conditions can be measured by
(for $r_{ph} > r_s$) \citep{2007ApJ...664L...1P}
\begin{equation}
    \mathcal{R}={{\left( \frac{F_{BB}}{\sigma T^{4}} \right)}^{1/2}},
\end{equation}
where $F_{tot}$ is the thermal emission flux and $\sigma$ is the Stefan - Boltzmann constant. The coasting values of the Lorentz factor is calculated as
\begin{equation}
    \Gamma ={{\left[ 1.06{{\left( 1+z \right)}^{2}}{{d}_{L}}\frac{Y{{\sigma }_{T}}{{F}_{BB}}}{2{{m}_{p}}{{c}^{3}}\mathcal{R}} \right]}^{1/4}}
\end{equation}
where $d_L$ is the luminosity distance, $\sigma_T$ and $m_p$ are the Thompson cross section and proton mass, respectively. $F_{BB}$ is the observed total flux of the BB component. The radius of the photosphere can be estimated by
\begin{equation}
    r_{ph}=\frac{{{L}_{0}}{{\sigma }_{T}}}{8\pi \Gamma _{ph}^{3}{{m}_{p}}{{c}^{3}}}
\end{equation}.
where $L_{0}=4\pi d^{2}_{L}YF^{obs}$ is the burst luminosity. The physical size at the base of the flow is
\begin{equation}
    {{r}_{0}}=\frac{{{4}^{3/2}}}{{{(1.48)}^{6}}{{(1.06)}^{4}}}\frac{{{d}_{L}}}{{{(1+z)}^{2}}}{{\left( \frac{{{F}_{BB}}}{Y{{F}_{tot}}} \right)}^{3/2}}\mathcal{R},
\end{equation}
where the parameter Y, related to the radiative efficiency of the bursts is defined as $Y = L_{0}/L_{obs}$ , $L_{obs}$ is the observed $\gamma$-ray luminosity. We set Y = 2 in our calculations.
\begin{equation}
r_{s}=\Gamma \times r_{0}
\end{equation}
The average values of the photosphere emission parameters of the three stages for the two GRBs are listed in Table 3. It is found that $\Gamma$ of the precursor and the extended emission are much smaller than that of the main emission for both bursts. The corresponding average values of the three phases of GRB 230307A are greater than those of GRB 211211A. Other corresponding parameters of the three stages for the two bursts are consistent. The evolution with time of the ratio $\mathcal{R}$, the initial radius $r_{0}$, the saturation radius $r_{s}$, and the photosphere radius $r_{ph}$ in the time-resolved spectra are demonstrated in Figures 18 and 19. The three radius almost keep constant with time. As shown in Figure 20, Lorentz factor $\Gamma$ varies with flux. This behaviour is similar to $E_{p}$ and $\alpha$ related to the rise and fall of flux. Therefore, the two GRBs show similar photosphere emission properties based on above analysis.

\begin{figure}[htbp]
\centering
\begin{minipage}[t] {\textwidth}
\centering
\includegraphics[width=\textwidth]{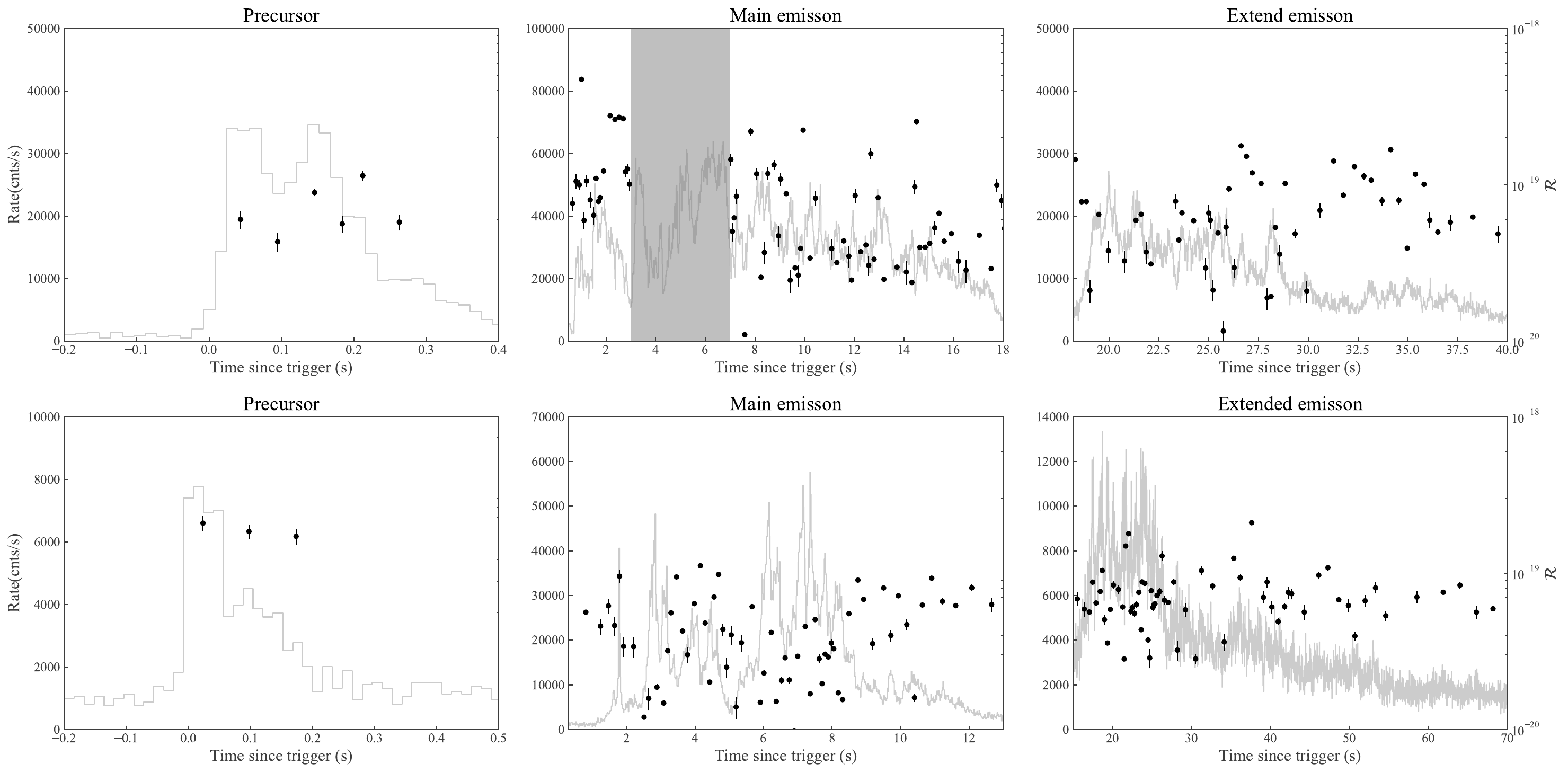}
\end{minipage}
\caption{Evolution of  $\mathcal{R}$ with time for the three phases.The upper three panels represent the precursor, main emission, extended emission of GRB 230307A, while lower three panels are the corresponding emission phase for GRB 211211A. \label{fig18}}
\end{figure}

\begin{figure}[htbp]
\centering
\begin{minipage}[t] {\textwidth}
\centering
\includegraphics[width=\textwidth]{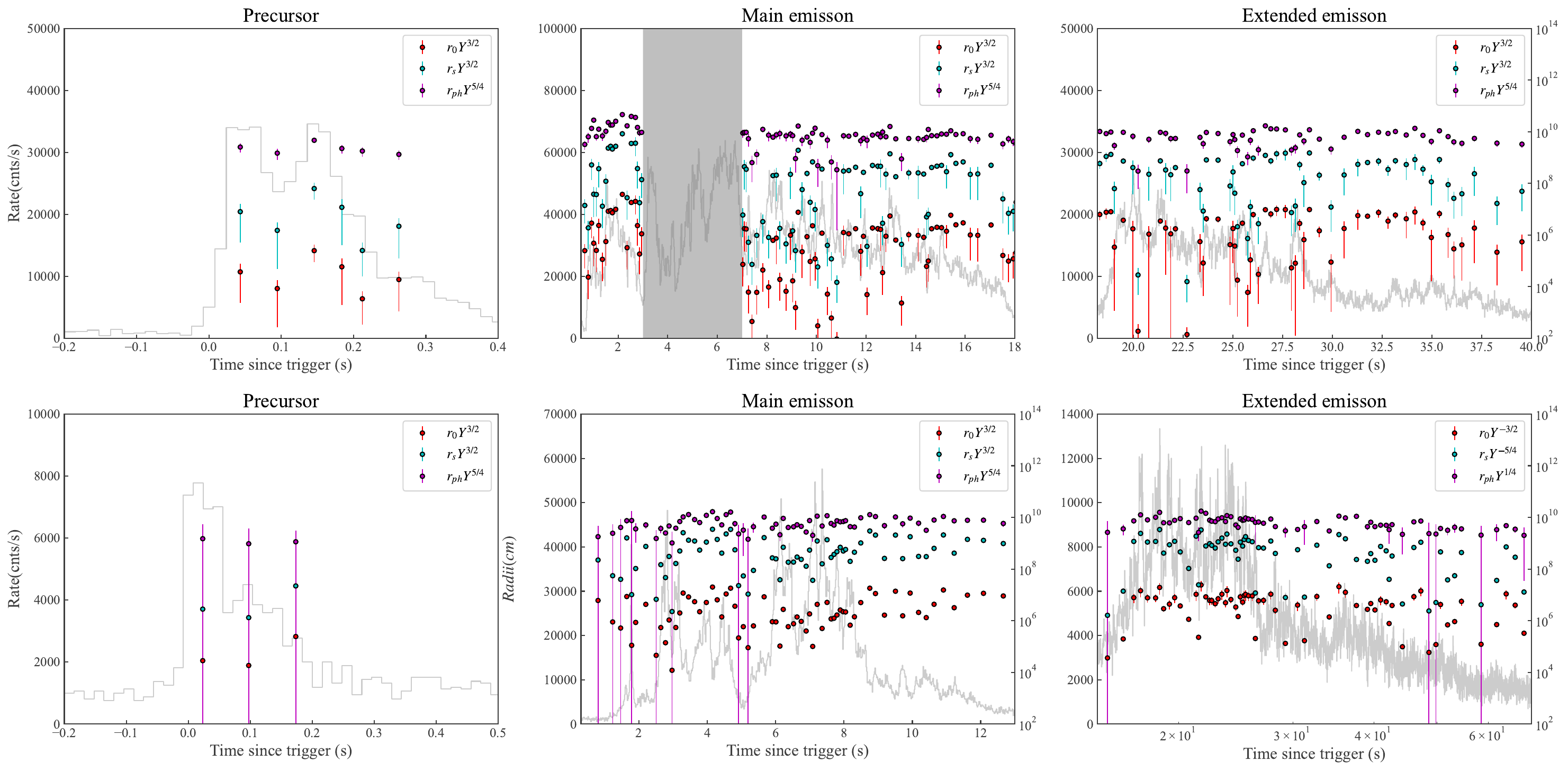}
\end{minipage}
\caption{Evolution of $r_{0}$, $r_{s}$ , and $r_{ph}$ with time for the three phases, where the red, green and the purple data points denote the $r_{0}$, $r_{s}$ , and $r_{ph}$, respectively. The upper three panels represent the precursor, main emission, extended emission of GRB 230307A, while lower three panels are the corresponding emission phase for GRB 211211A. \label{fig19}}
\end{figure}

\begin{figure}[htbp]
\centering
\begin{minipage}[t] {\textwidth}
\centering
\includegraphics[width=\textwidth]{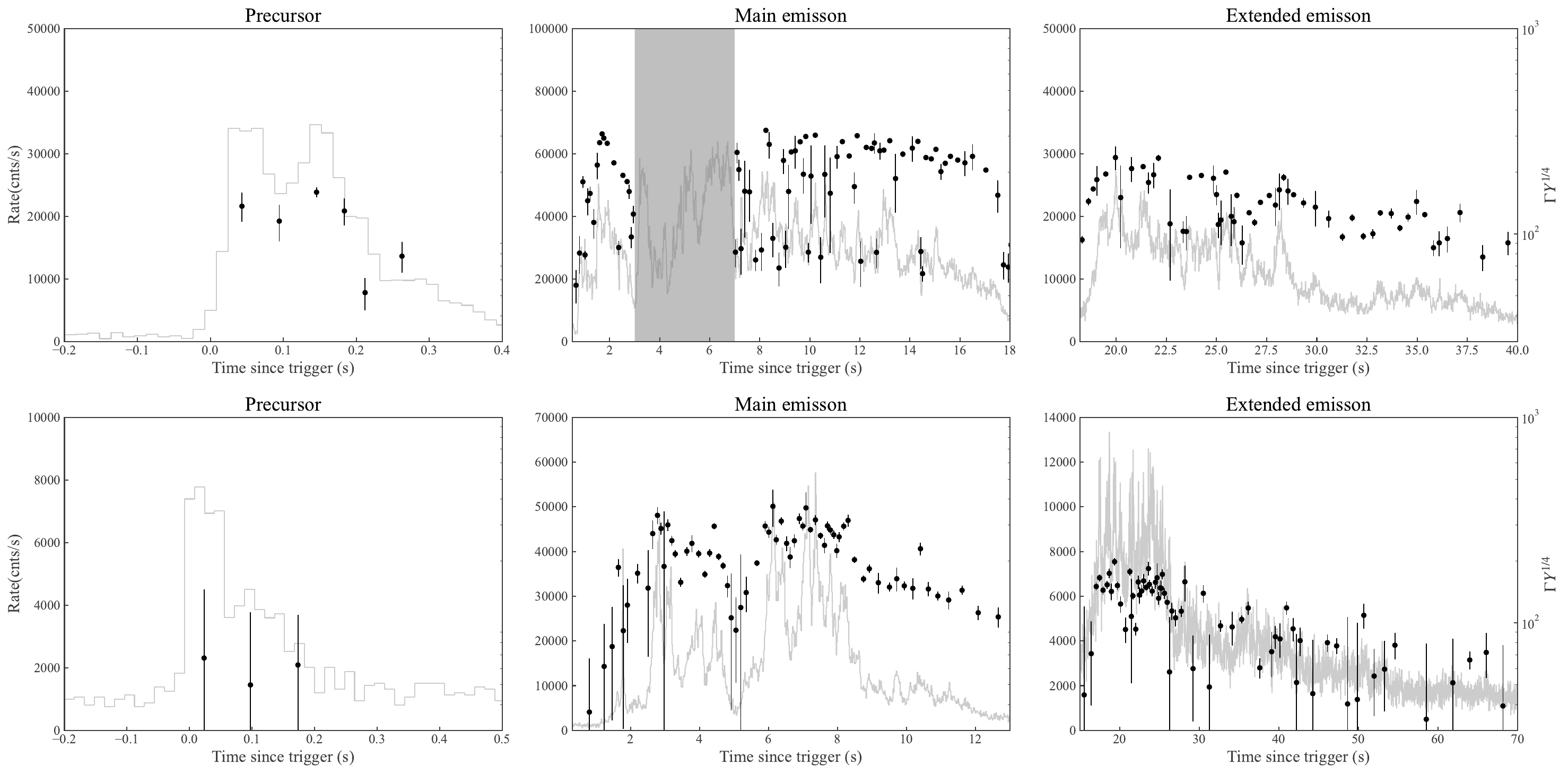}
\end{minipage}
\caption{Evolution of $\Gamma$ with time for the three phases. The upper three panels represent the precursor, main emission, extended emission of GRB 230307A, while lower three panels are the corresponding emission phase for GRB 211211A. \label{fig20}}
\end{figure}

\begin{deluxetable}{cccccccc}
\tablecaption{Photosphere emission parameters comparison between GRB 211211A and GRB 230307A}
\tablehead{
\colhead{}  & \colhead{Precursor} & \colhead{Main emission}  &\colhead{Extended emission}  
}
 \startdata
\hline
\multicolumn{4}{c}{GRB 230307A}\\
\hline
R&$(7.00\pm2.46)\times 10^{-20}$&$(7.56\pm 7.30)\times10^{-20}$&$(7.79\pm 3.89)\times10^{-20}$ \\
$\Gamma Y^{1/4}$ &$257.48\pm49.90$ &$350.71\pm124.70$&$206.66\pm55.26$\\
$r_0Y^{-3/2}(cm)$&$(3.45\pm4.30)\times10^{5}$&$(2.47\pm4.82)\times10^6$&$(3.48\pm3.28)\times10^6$\\
$r_sY^{-5/4}(cm)$&$(9.63\pm12.12)\times10^{7}$&$(7.22\pm13.31)\times10^8$&$(6.37\pm6.02)\times10^8$\\
$r_{phY^{1/4}}(cm)$&$(1.30\pm0.36)\times10^{10}$&$(1.63\pm1.29)\times10^{10}$&$(1.08\pm0.41)\times10^{10}$\\
\hline
\multicolumn{4}{c}{GRB 211211A}\\
\hline
R&$(5.90\pm4.30)\times 10^{-20}$&$(4.13\pm 2.61)\times10^{-20}$&$(7.31\pm 3.18)\times10^{-20}$ \\
$\Gamma Y^{1/4}$&$131.92\pm23.82$ &$319.13\pm124.36$&$160.23\pm42.89$\\
$r_0Y^{-3/2}(cm)$&$(5.24\pm5.32)\times10^{5}$&$(4.12\pm5.04)\times10^6$&$(6.02\pm5.21)\times10^6$\\
$r_sY^{-5/4}(cm)$&$(5.97\pm5.34)\times10^{7}$&$(9.93\pm10.21)\times10^8$&$(1.00\pm0.91)\times10^9$\\
$r_{phY^{-1/4}}(cm)$&$(6.81\pm1.71)\times10^{9}$&$(0.39\pm4.25)\times10^{9}$&$(9.73\pm3.39)\times10^{9}$\\
\hline
\enddata
\end{deluxetable}

\section{Discussion and Conclusions\label{Summary}}
GRB 230307A and GRB 211211A are two peculiar LGRBs with some SGRB features. In this paper, we have analyzed in detail the temporal and spectral data of the two bursts observed by Fermi-GBM detector. Different from other work, we systematically compare the temporal and spectral properties of the two GRBs to explore the common properties of the two GRBs. 

We find that both GRB 230307A and GRB 211211A have three stages in the lightcurves, that is, a dimmer precursor, a much brighter main emission and an extended emission. LGRBs usually have significant spectral lags. In this paper, we find the spectral lags of the three phases in the two bursts are nearly zero. Zero spectral lag is often observed in SGRBs (e.g. \cite{2006MNRAS.367.1751Y}). Thus, the tiny spectral lags put the two bursts into SGRB population rather than LGRB population.

GRB lightcuve is highly variable, and MVT can be adopted to study the GRB lightcurve variability. The MVT of long bursts with a typical value of 200 ms is much larger than that of short ones with a typical value of 10 ms. The very short MVT numbers (39 ms for GRB 230307A and 49 ms for GRB 211211A) indicate that the two GRBs might belong to the short burst population. The three phases with tiny MVT values for the two bursts also should belong to short bursts. When considering MVT and the duration simultaneously GRB230307A lies very close to the GRB 211211A  (solid circle and solid pentagrams Figure 3) and both the two GRBs lie in the region populated by other short bursts with the extended emission (SEE). Figure 3 indicates that the extended emission can be a kind of feature for long-lasting merger. Our results are consistent with other investigations \citep{2023GCN.33577....1C}. We expect that the two GRBs share the same merger origin. 

Through the time-integrated spectral analysis of the three phases and the whole emission we can compare the Amati relation of the two GRBs, respectively. It is found that both the precursors seem to evidently lie in the SGRB region in the Amati relation plane, and the whole emission of the two GRBs lie closer to the track of LGRBs. 

Similar to the Amati relation the hardness-duration relation of the two GRBs reveals a LGRBs origin for the whole emission including the main and extended emission as shown in hardness-duration plane. While the precursors appear to obviously lie in the region of SGRBs. We suspect that the precursor seems to share the properties of the short bursts based on spectral lag, MVT, the Amati and hardness-duration relation of the two GRBs, which deserves a further investigation with greater sample. 
  
Other observations from GECAM and LEIA draw the same conclusion that the two GRBs are type I GRBs and have a compact star merger origin. \cite{2023GCN.33569....1L} performed the observation of GRB 230307A with James Webb Space Telescope and suggested that a kilonova origin is the most likely interpretation due to the very red colour and the absence of a host galaxy. The observation also implies that GRB 230307A also originates from a binary compact star merger.

Our analysis of the the spectral widths of the two bursts show that their spectra are wider than those of normal SGRBs, which seems to reveal that the two GRBs are LGRBs. To compare the two GRBs in detail if they belong to the short or long class we list all the methods we tested in Table 4. Based on above analysis and the results of Table 4, it seems that the short and long GRB classifications and the related physical origins are complex, and it is hard to draw a solid conclusion on this issue.   

Through the time-resolved spectral analysis of the two GRBs, we find that the modelling parameters evolved with time share the same evolutionary trend with the burst flux. The "double tracking" model for both the peak energy and the low-energy spectral index tracking the flux further support the same spectral evolution. The corresponding photosphere emission properties of the two GRBs are also shown in this paper.

The physical mechanism of GRB prompt emission is still an unresolved issue, and the time-resolved spectral analysis can provide some evidences to investigate radiation mechanism. Due to the very high brightness of the two bursts, we can obtain the time-resolved spectral data of the three stages with high signal-noise ratio. It is found that almost all the time-resolved spectra are quasi-thermal.

It is well known that the photospheric model has been suggested in GRB research field. The sub-photosphere turns to be transparent, and the thermal component is shown in the precursor spectrum (e.g., \cite{2000ApJ...530..292M,2006ApJ...642..995P,2013ApJ...777...62I}).  However, \citet{2005MNRAS.357..722L} tested GRB precursor spectra. It seems that a power-law model, instead of a blackbody model, is preferred for the spectral data. \cite{2021ApJS..252...16L} studied the temporal properties of the precursors in SGRBs. They proposed that the precursor, the main burst, and the extended emission have same physical origin. \cite{2022A&A...657A.124L} studied the precursors of LGRBs, and the temporal behavior of the precursor is similar to that of the main burst. In our work, it is shown that the precursor, the main emission, and the extended emission are dominated by the thermal emission. It is reasonable to apply the photosphere model to explain the origins of the precursor and main emission for the two GRBs.
 
GRB jet structure is one topic that is widely discussed. Cocoon is suggested to be interacted with GRB jet \citep{2005ApJ...629..903L}. This scenario was applied to the case of GRB 170817A \citep{2019ApJ...881...89L,2020ApJ...898...59L}. The cocoon emission may have thermal component \citep{2018MNRAS.473..576G}. In such case, although GRB precursor may not be affected by the cocoon by a direct way, the jet with the propagation collimated by the cocoon may have strong thermal emission.  SGRB with extended emission has several kinds of explanations. Magnetar-related properties have been suggested \citep{2014MNRAS.438..240G}. We may consider the fallback related to the accretion \citep{2007MNRAS.376L..48R}. In order to interpret the thermal component shown in the extended emission, the radiation from the fallback due to accretion sounds interesting. In addition, the X-ray emission of some GRBs have thermal component (Straling et al. 2012; Bellm et al. 2014). Bremsstrahlung can be a possible mechanism for the thermal component (Liu \& Mao 2019). However, we cannot apply an unified model to fully explain the thermal and non-thermal spectral properties of the two bursts. The observational founding in this paper adds an challenge in GRB physics. 

The observed spectral widths of both bursts evolve from narrow to wide with time. It is indicated that the GRB jet may have transition from fireball to Poynting-flux. However, the detail time-resolved spectral analysis reveals that the two bursts are quasi-thermal-dominated spectra across the whole emission phase rather than following the evolution from thermal to non-thermal. Further investigation show that the contributions of thermal flux to the total flux unchangeably. Therefore, the quasi-thermal radiations of the two bursts do not support the transition of their jet from fireball to Poynting-flux. Moreover, the spectral widths of the two bursts are much wider than those of other SGRBs. We note that the spectral widths measured by the spectral fitting of empirical models could not be a reliable tool to infer physical properties on the GRB emission \citep{2019A&A...629A..69B}. Further examinations of spectral fitting should be applied in the future to physically reveal the GRB radiation processes.    

In summary, the consistent temporal and spectral properties of the two bursts revealed in this paper suggest the GRB 230307A and GRB 211211A are the very similar bursts, which is very rare thing in GRB field. 

\begin{deluxetable}{cccccccc}
\tablecaption{A list of the method we adopted and the GRB classification for the GRB 230307A and GRB 211211A}
\tablehead{
\colhead{stage}  & \colhead{GRBname} & \colhead{lag}  &\colhead{MVT} &\colhead{MVT-T90}  &  \colhead{Amati relation}  &\colhead{HR-T90}  & \colhead{spectral width} 
}
 \startdata
 \multirow{2}{*}{Precursor}&GRB 230307A& S & S & S  & S & S &  \\
&GRB 211211A &S & S & S  & S & S & \\
\hline
\multirow{2}{*}{Main emission}&GRB 230307A& S & S & S  & T& L &\\
&GRB 211211A & S & S & S  & T & L &\\
\hline
\multirow{2}{*}{Extended emission}&GRB 230307A& S & S & S  & T & L&\\
&GRB 211211A & S & S & S & L & L&\\
\hline
\multirow{2}{*}{Whole emission}&GRB 230307A&S & S & S  & L & L & L\\
&GRB 211211A & S & S & S  & L & L& L \\
\hline
\enddata
\tablecomments{The "S", "L", and "T" denotes the SGRB, LGRB, and the transition GRB, respectively.}
\end{deluxetable}

\begin{acknowledgments}
\indent We acknowledge the use of the public data from the Fermi data archives. This work has the financial support of the National Key R\&D Program of China (2023YFE0101200), the National Natural Science Foundation of China (grant 12163007, 11673062), Key Laboratory of Colleges and Universities in Yunnan Province for High-energy Astrophysics, National Astronomical Observatories Yunnan Normal University Astronomical Science Popularization Education Base. JM is supported by the National Natural Science Foundation of China 12393813, CSST grant CMS-CSST-2021-A06, and the Yunnan Revitalization Talent Support Program (YunLing Scholar Project), 
\end{acknowledgments}

\bibliography{paperR2}{}
\bibliographystyle{aasjournal}
\appendix
\startlongtable
\begin{deluxetable}{cccccccccc}
\tablecaption{Time-resolved spectral fitting result of GRB 230307A and GRB 211211A.\label{}}
\tablehead{
\colhead{$t_{start}-t_{end}$} & \colhead{$S$} & \colhead{Model} & \colhead{$\alpha$} & \colhead{$\beta$} & \colhead{$E_p$} & \colhead{kT} &\colhead{$F_{bb} \times 10^{-6}$}&\colhead{$F\times 10^{-6}$}& \colhead{DIC}\\
\colhead{s}&\colhead{} &\colhead{} &\colhead{} &\colhead{} &\colhead{$keV$} &\colhead{$keV$}&\colhead{$erg^{-1}cm^{-2}s^{-1}$}& \colhead{$erg^{-1}cm^{-2}s^{-1}$} &\colhead{}\\
\colhead{(1)}&\colhead{(2)}&\colhead{(3)}&\colhead{(4)}&\colhead{(5)}&\colhead{(6)}&\colhead{(7)}&\colhead{(8)}&\colhead{(9)}&\colhead{(10)}\\
}
\startdata   
\hline
 &   &  &  &  &GRB 230307A  &  &  & &  \\
\hline
\multirow{4}{*}{0.02$\sim$0.07}&\multirow{4}{*}{83}&CPL&${-0.55}^{+0.06}_{-0.06}$&...&${192.02}^{+5.89}_{-6.66}$&...&...&${6.54}^{+6.35}_{-6.72}$&121.28\\
&&Band&${-0.45}^{+0.08}_{-0.09}$&${-3.33}^{+0.2}_{-0.21}$&${175.1}^{+9.24}_{-8.38}$&...&...&${6.57}^{+0.19}_{-0.19}$&128.73\\
&&CPL+BB&${-0.48}^{+0.13}_{-0.0}$&...&${176.72}^{+3.99}_{-18.26}$&${163.01}^{+104.33}_{-6.93}$&${2.61}^{+2.37}_{-2.81}$&${6.54}^{+6.35}_{-6.72}$&121.28\\
&&Band+BB&${-0.45}^{+0.08}_{-0.09}$&${-3.33}^{+0.2}_{-0.21}$&${175.1}^{+9.24}_{-8.38}$&${89.88}^{+45.51}_{-31.67}$&${0.46}^{+0.42}_{-1.42}$&${6.57}^{+0.19}_{-0.19}$&128.73\\
\hline
\enddata
\tablecomments{Table 5 is published in its entirety in the machine-readable format. A portion is shown here for guidance regarding its form and content.}
\end{deluxetable}

\end{document}